\documentclass[acmtog, nonacm]{acmart}
\acmSubmissionID{1445}

\usepackage{textcomp} 

\newcommand{\sysname}{\textcolor{black}{\textit{WorldPrompter}}\xspace}

\newcommand{\new}[1]{\textcolor{black}{#1}}

\newcommand{\rev}[1]{\textcolor{black}{#1}}

\newcommand{\figref}[1]{Fig.~\ref{#1}}

\usepackage[ruled]{algorithm2e} 

\SetAlFnt{\small}
\SetAlCapFnt{\small}
\SetAlCapNameFnt{\small}
\SetAlCapHSkip{0pt}

\acmJournal{TOG}

\begin{document}
\title{Generating 360° Video is What You Need For a 3D Scene}


\author{Zhaoyang Zhang}
\orcid{0000-0002-8892-1191}
\authornote{Work partially done during internship at Adobe Research.}
\affiliation{%
 \institution{Yale University}
 \streetaddress{51 Prospect St.}
 \city{New Haven}
 \state{CT}
 \postcode{06510}
 \country{USA}
}
\affiliation{%
 \institution{Adobe Research}
 \streetaddress{345 Park Ave.}
 \city{San Jose}
 \state{CA}
 \postcode{95110}
 \country{USA}
}
\email{zhaoyang.zhang@yale.edu}

\author{Yannick Hold-Geoffroy}
\orcid{0000-0002-1060-6941}
\affiliation{%
 \institution{Adobe Research}
 \streetaddress{345 Park Ave.}
 \city{San Jose}
 \state{CA}
 \postcode{95110}
 \country{USA}
}
\email{holdgeof@adobe.com}

\author{Milo\v{s} Ha\v{s}an}
\orcid{0000-0003-3808-6092}
\affiliation{%
 \institution{Adobe Research}
 \streetaddress{345 Park Ave.}
 \city{San Jose}
 \state{CA}
 \postcode{95110}
 \country{USA}
}
\email{milos.hasan@gmail.com}

\author{Ziwen Chen}
\orcid{0009-0007-1776-7696}
\affiliation{%
 \institution{Oregon State University}
 \streetaddress{2500 NW Monroe Ave.}
 \city{Corvallis}
 \state{OR}
 \postcode{97331}
 \country{USA}
}
\email{chenziw@oregonstate.edu}

\author{Fujun Luan}
\orcid{0000-0001-5926-6266}
\affiliation{%
 \institution{Adobe Research}
 \streetaddress{345 Park Ave.}
 \city{San Jose}
 \state{CA}
 \postcode{95110}
 \country{USA}
}
\email{luanfj11@gmail.com}

\author{Julie Dorsey}
\orcid{0000-0003-2495-4979}
\affiliation{%
 \institution{Yale University}
 \streetaddress{51 Prospect St.}
 \city{New Haven}
 \state{CT}
 \postcode{06510}
 \country{USA}
}
\email{julie.dorsey@yale.edu}

\author{Yiwei Hu}
\orcid{0000-0002-3674-295X}
\affiliation{%
 \institution{Adobe Research}
 \streetaddress{345 Park Ave.}
 \city{San Jose}
 \state{CA}
 \postcode{95110}
 \country{USA}
}
\email{yiwhu@adobe.com}

\begin{abstract}
\new{Generating 3D scenes is still a challenging task due to the lack of readily available scene data. Most existing methods only produce partial scenes and provide limited navigational freedom. 
We introduce a practical and scalable solution that uses 360° video as an intermediate scene representation, capturing the full-scene context and ensuring consistent visual content throughout the generation.} We propose \sysname, a generative pipeline that synthesizes traversable 3D scenes from text prompts. \sysname incorporates a conditional 360° panoramic video generator, capable of producing a 128-frame video that simulates a person walking through and capturing a virtual environment. The resulting video is then reconstructed as Gaussian splats by a fast feedforward 3D reconstructor, enabling a true walkable experience within the 3D scene. \new{Experiments demonstrate that our panoramic video generation model, trained with a mix of image and video data, achieves convincing spatial and temporal consistency for static scenes. This is validated by an average COLMAP matching rate of 94.6\%, allowing for high-quality panoramic Gaussian splat reconstruction and improved navigation throughout the scene.} Qualitative and quantitative results also show it outperforms the state-of-the-art 360° video generators and 3D scene generation models. 
\end{abstract}

%
%


%
%


\begin{teaserfigure}
\centering
\begin{figure}[H]
    \centering
    \includegraphics[width=\textwidth]{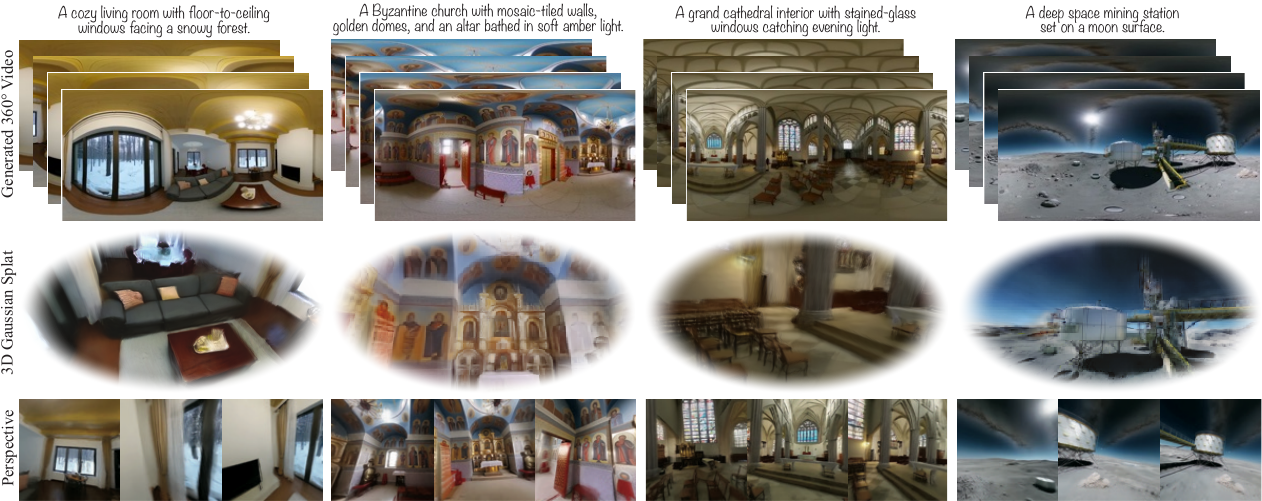}
    \caption{We present \sysname, a method that generates traversable environments from a text prompt alone (above). Our method first generates a spatially coherent 360° video capture (top) that travels through a static environment described by the prompt. Next, we elevate generated video capture into a 3D Gaussian Splat (middle), enabling free navigation within a 3D scene using a virtual perspective camera to generate novel views (bottom).}
    \label{fig:teaser}
\end{figure}
\end{teaserfigure}

\maketitle

\section{Introduction} \label{sec:intro}
Scene-level 3D generation has attracted increasing research interest in recent years. \new{The limited availability of data generally hinders direct 3D scene generation \cite{liu2024pyramiddiffusionfine3d}, making it difficult for methods to generalize to realistic scene synthesis effectively. To address this challenge, recent approaches typically rely on an image diffusion model to create realistic scene views.} For example, methods like RealmDreamer \cite{shriram2024realmdreamer} and LucidDreamer \cite{chung2023luciddreamer} can generate Gaussian Splatting (GS) 3D scenes from text prompts. They rely on single-view diffusion models and monocular depth estimation to generate scenes progressively from one view to another by 2D inpainting. However, the 3D scenes created by these methods are confined to a limited viewable area and are not freely traversable. This constraint differentiates these approaches from a "complete" scene generation that offers a navigable virtual world, similar to what is found in interactive video games. \new{Recent works \cite{zhou2025dreamscene360, yang2024layerpano3d} attempt to bridge the gap between partial and full scene generation by leveraging panoramic images. However, these methods still rely on outpainting to fill in missing regions, which inherently limits the navigable range and often leads to distortions during viewpoint transitions. }

\new{More importantly, we observe that scenes generated via these outpainting-based approaches frequently suffer from distortion and inconsistencies in visual content across the scene. We hypothesize that these issues stem from the commonly used progressive outpainting approach, which independently synthesizes new content based on the previously generated content. However, this assumption is flawed, as visual content is globally interconnected and consistent rather than strictly sequential \cite{tian2024VAM}. As a result, while their generated scenes may appear locally plausible, they exhibit significant global distortions (Fig. \ref{fig:scene_gen_comparison}).}

In this paper, \new{instead of synthesizing a scene in a view-by-view fashion, we aim to generate a maximally navigable scene in a single pass, ensuring undistorted global visual content synthesis.} 
\new{Although scenes can be represented in various formats, such as meshes or point clouds, data in these formats is often limited and difficult to scale.} Drawing inspiration from conventional 3D capture methods and recent advancements in video generation models (\cite{polyak2024moviegencastmedia, opensora, blattmann2023stablevideodiffusionscaling}), we propose using videos as an intermediate representation to model complete 3D environments. \new{Modeling scenes using video captures provides a scalable approach for creating training data, enabling generative models to achieve strong generalization capabilities.} To validate this idea, we propose \sysname, a generative pipeline that produces a navigable 3D scene from text prompts. Specifically, we train a diffusion model to generate videos traversing the environment based on text prompts. Moreover, instead of synthesizing captures from perspective cameras, we leverage panoramic cameras that directly capture the entire surroundings of an environment, providing 360° viewpoints that are ideal for our application. 
 
 \sysname is a two-stage pipeline composed of a 360° panoramic video generation model and a 3D reconstruction step to generate the final 3D Gaussians, a summary of which is shown in \figref{fig:teaser}. The text-conditioned panoramic video generation model is trained to generate a 360° panoramic video that simulates a person walking through an environment while holding a panoramic camera to capture the scene. To train the video generator, we collected a dataset comprising approximately 1,700 panoramic videos and a panorama image dataset containing more than 200,000 high-quality panoramic in-the-wild captures. To enhance the generative capability of our method, we fine-tuned a Diffusion Transformer (DiT)-based video generator on our dataset to produce 10.6-second 360° panoramic videos (128 frames at 12 fps). We introduce a masked diffusion loss to ensure a clean panoramic capture sequence free from the person holding the camera, ready for subsequent reconstruction purposes. From the generated videos, we extract panoramic frames and project them from their original spherical coordinates onto the image plane to obtain perspective images from multiple viewpoints throughout the scene. The reconstruction step then builds the final 3D Gaussian Splat (3DGS) scene from these cropped images using a fast feedforward reconstruction model \cite{ziwen2024llrm}. The resulting scene allows users to navigate freely within a virtual environment, \new{beyond the original video camera path.} 

\new{We demonstrate that our 360° video generation model---trained on a mixture of image and video data---achieves state-of-the-art performance in temporal consistency and visual quality.} \new{The high consistency of the views, confirmed by a matching rate of up to 94.5\% using COLMAP \cite{schoenberger2016sfm}, ensures that the videos can be reliably used as proxies for 3D scene generation. This results in a traversable, 360° viewable environment with globally coherent visual content. We demonstrate that our generated scene allows longer navigation range and offers superior global visual coherence and quality compared to prior art.} We highlight our contributions as follows:
\begin{itemize}
    \item \new{We propose leveraging panoramic capture videos as an intermediate 3D scene representation. We found it is a practical solution for scaling up training data for high-quality photorealistic scene synthesis.}
    \item \new{We present \sysname, a practical generative text-to-scene pipeline that synthesizes a traversable 3D scene with strong global visual coherence.}
    \item \new{We introduce a high-quality 360° panoramic video generation model, where the proposed mixed datasets and masked diffusion loss play a critical role in achieving artifact-free and generalizable static 360° scene generation.}
\end{itemize} 
\begin{figure*}[ht]
    \centering
    \includegraphics[width=0.99\textwidth]{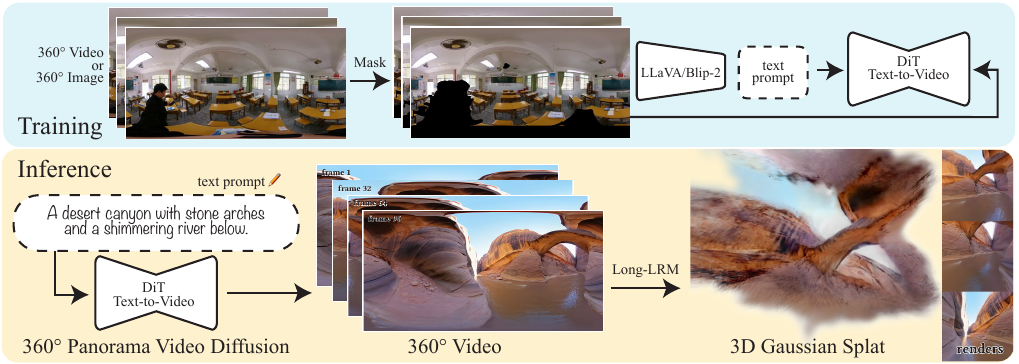}
    \caption{\textbf{Overview of \sysname{}}. (top) We train a text-to-video model on a mix of 360° videos and images depicting in-the-wild environments. As it is challenging to avoid the person and camera equipment being visible in the video capture, we mask these elements out of the frame using a pretrained image segmentation model \cite{SelectSuject}, and obtain the prompts from the video frames using LLaVA \cite{liu2023llava} for the videos, and BLIP-2 \cite{li2023blip} for images. (bottom) At inference time, a user supplies a text prompt to our text-to-video model, which produces a ``walk-through'' video of the scene, which we reconstruct into a 3D Gaussian splat representation using Long-LRM \cite{ziwen2024llrm}.} 
    \label{fig:pipeline}
\end{figure*}

\section{Related Work}

\subsection{Panoramic Image and Video Generation}
In computer graphics, 360° panoramas---often called IBLs or environment maps---are widely used to represent distant lighting and environments \cite{debevec2023recovering}. Numerous studies estimate lighting as High Dynamic Range (HDR) panoramas from input photos \cite{zhan2021emlight, hweberEditableIndoorLight, dastjerdi2023everlight}. However, the resolution of these panoramas is often limited, making them suitable for light representation but insufficient for 3D reconstruction. 

Advancements in diffusion models have made high-resolution image synthesis possible \cite{rombach2022highresolutionimagesynthesislatent, esser2024scaling}. Panorama generation also benefits from the use of diffusion models. For example, Text2Light \cite{chen2022text2light} generates 4K HDR panoramas via text-driven low-resolution generation and super-resolution inverse tone mapping. MVDiffusion \cite{Tang2023mvdiffusion} creates panoramas by stitching consistent multi-view images generated from text prompts using attention layers. PanFusion \cite{panfusion2024} employs a dual-branch diffusion model with equirectangular-Perspective Projection Attention (EPPA) to improve consistency and control across panorama and perspective domains. \rev{DiffPano \cite{ye2024diffpanoscalableconsistenttext} fine-tunes Stable Diffusion \cite{rombach2022highresolutionimagesynthesislatent} using LoRA \cite{hu2021loralowrankadaptationlarge} to generate panorama images, and with a Spherical Epipolar-Aware Attention Module, it enables multi-view panorama generation.}

Diffusion models have also demonstrated their potential in video generation \cite{blattmann2023align, blattmann2023stablevideodiffusionscaling}. Diffusion Transformers \cite{peebles2023scalable}  offer a more versatile method for modeling high-dimensional data, making them particularly well-suited for video generation \cite{videoworldsimulators2024}. For 360° video generation, \citeauthor{wang2024360dvd} propose 360DVD for generating 360° panoramic videos from text prompts and motion conditions, where they leverage a lightweight 360-Adapter to finetune the Stable Diffusion \cite{rombach2022highresolutionimagesynthesislatent} text-to-image model. However, their model is limited to generating only 16 frames, whereas our 360° video generation model produces significantly more frames with superior temporal consistency.

\subsection{3D Neural Representation and Reconstruction}
3D reconstruction and representation is an extensively studied field. While mesh-based methods \cite{kazhdan2006poisson} and point-based methods \cite{qi2017pointnet} have been extensively explored, they are limited in the quality of their results. In contrast, neural representations offer a flexible way to reconstruct 3D data. For instance, Neural Radiance Fields (NeRF) \cite{mildenhall2020nerf} introduces a novel way to represent 3D scenes by encoding volumetric scene information as a continuous function parameterized by a neural network, which inspired many follow-up works \cite{mueller2022instant, barron2022mipnerf360unboundedantialiased, barron2023zipnerfantialiasedgridbasedneural} that optimize efficiency, quality, and rendering speed. However, the MLP-based paradigm requires training a distinct model per scene, making reconstruction time-consuming. Additionally, rendering views involves model queries, creating a bottleneck for real-time applications. 
3D Gaussian Splatting (3DGS) \cite{kerbl3Dgaussians} introduces a memory-efficient and low-overhead 3D implicit representation using Gaussians derived from Structure-from-Motion (SfM) points. By combining anisotropic covariance optimization with a fast visibility-aware rendering algorithm, 3DGS enables efficient scene representation, high-quality real-time rendering, and has quickly emerged as a new paradigm in neural reconstruction \cite{ye2024gsplatopensourcelibrarygaussian, wu2024recentadvances3dgaussian, COGS2024, MeshGaussian2024, charatan2024pixelsplat3dgaussiansplats}. Nevertheless, 3DGS still requires an optimization process, which can be costly for larger scenes. Recent advances such as Large Reconstruction Models (LRMs) \cite{ziwen2024llrm, wei2024meshlrm, xie2024lrmzerotraininglargereconstruction, gslrm2024, wang2023pf} target scene reconstruction with a single feed-forward evaluation during inference. They achieve a reconstruction quality comparable to optimization-based methods while significantly improving their efficiency. In our scene generation pipeline, we also utilize this state-of-the-art feed-forward reconstruction approach for long-sequence reconstruction, achieving fast reconstruction. 

\subsection{3D Scene Generation}
3D generation has emerged as a prominent focus in generative modeling. Given the scarcity of high-quality 3D data, recent works usually leverage intermediate representations, such as multi-view images \rev{\cite{liu2023zero1to3zeroshotimage3d, voleti2024sv3dnovelmultiviewsynthesis, Hu_2024_CVPR, li2023instant3dfasttextto3dsparseview, shi2024mvdreammultiviewdiffusion3d, liu2023one2345} and videos \cite{parthasarathy2024vid3dsynthesisdynamic3d, han2024vfusion3dlearningscalable3d, chen2024v3dvideodiffusionmodels}}, to improve the quality and efficiency of their reconstructions. Many studies have explored generating 3D objects in various formats, including Gaussian splats and meshes. For instance, Instant3D \cite{li2023instant3dfasttextto3dsparseview} accelerates 3D asset creation from text prompts by combining sparse-view generation with transformer-based reconstruction. Similarly, MVDream \cite{shi2024mvdreammultiviewdiffusion3d} improves consistency through multi-view diffusion by integrating 2D and 3D data. One-2-3-45 \cite{liu2023one2345} simplifies single-image 3D reconstruction, enabling view-consistent mesh generation. 

Scene-level generation, however, is a more challenging task because it requires more observations to achieve comprehensive view coverage. Typically, 3D scene generation methods employ progressive pipelines \cite{ouyang2023text, chung2023luciddreamer, hollein2023text2room}: anchor cameras are first selected to generate text-aligned 2D images and corresponding depth maps, followed by a neural reconstruction process to build the initial 3D structure. This process is iteratively repeated to expand scene coverage by generating additional camera views and inpainting missing regions. However, such methods often suffer from prolonged generation times and incomplete coverage. Other studies have explored interactive and layout-guided approaches for 3D scene generation. Systems like WonderWorld \cite{yu2024wonderworldinteractive3dscene} provide user-friendly interfaces for scene customization, while Director3D \cite{li2024director3d} facilitates real-world scene creation with adaptive camera trajectories. \new{Another potential limitation of this family of methods, as discussed in Sec. \ref{sec:intro}, is the lack of guaranteed global visual coherence, often resulting in distorted scene layouts and object shapes.} \rev{Most recent literature has discovered the potential of video generation for consistent scene-level generation. VideoScene \cite{wang2025videoscenedistillingvideodiffusion} leverages the temporal consistency in video diffusion models to generate novel views for 3D reconstruction by interpolating two input keyframes. \citeauthor{yu2024viewcraftertamingvideodiffusion} and \citeauthor{wu2025videoworldmodelslongterm} explore explicitly reconstructing a 3D point cloud as spatial memory for consistent video generation.} 

The closest method to ours is DreamScene360 \cite{zhou2025dreamscene360}, which generates 3D scenes from text prompts using a single panorama image, along with depth estimation and Gaussian Splats optimization. However, while the resulting scene supports 360° viewing, it suffers from a limited movable range. \new{LayerPano3D \cite{yang2024layerpano3d} offers slightly increased navigational range by generating layered panoramas, but the content in the additional layers may suffer from distortion.} ODIN \cite{wallingford2024image} proposed learning 3D from 360° videos. However, their approach extracts only cropped views from videos to perform image-to-image style novel view synthesis, underutilizing the full potential of 360° videos. As a result, ODIN fails to model a complete 3D world and can only generate partial scenes. Beyond 3D, 4K4DGen \cite{li20244k4dgenpanoramic4dgeneration}, introduces panorama-to-4D generation but still remains limited in terms of traversable area. 

\new{In contrast to these approaches, our model introduces 360° videos as an explicit 3D scene proxy in the generation task, offering greater data scalability and ensuring strong visual coherence alongside comprehensive view coverage, ultimately enabling a realistic navigation experience.}

\section{\textit{WorldPrompter}}

\subsection{Overview}
\new{We show that 360° video capture serves as an effective, all-you-need proxy for 3D scene generation, as validated by the \sysname{} pipeline.} It synthesizes a traversable, 360° viewable 3D scene in two stages: \textit{generation} and \textit{reconstruction}. The \textit{generation} stage leverages a text-conditioned 360° panoramic video diffusion model to generate a 10.6-second panoramic capture video for a target scene, simulating someone holding the camera and walking through the environment. Later, the \textit{reconstruction} stage runs COLMAP \cite{schoenberger2016sfm} to estimate precise camera poses from perspective crops of the generated videos and builds a navigable 3DGS using Long-LRM \cite{ziwen2024llrm} with the calibrated camera poses. The generated panoramic video features rich details and geometrically consistent visual details \new{over the entire scene}, ensuring high-quality reconstruction and yielding appealing novel views in the traversable scene.

\subsection{Video Generator for 360° Panoramic Captures}
Training a generative model for 360° videos from scratch is challenging; therefore, we start with an existing video generator. 
\rev{We leverage a pretrained text-to-video diffusion model as backbone, and fine-tune it using panoramic data. }

Specifically, our model's backbone architecture $f_{\theta}$ is based on Diffusion Transformers \cite{peebles2023scalable}. Typically, in latent DiT models, a video or an image is first encoded by a 3DVAE \cite{yu2024language}, resulting a latent visual sequence $x \in \mathbb{R}^{T \times 3 \times H \times W}$ ($T=1$ if it's an image). The visual sequence is tokenized into patches, yielding a set of visual tokens $\hat{x} \in \mathbb{R}^{K \times D}$, where $T$, $H$, and $W$ correspond to the temporal and spatial dimensions of the video, $K$ represents the total number of tokens, and $D$ denotes the feature dimension. Being a text-conditioned model, it requires text prompts $c$ that we encode using a pre-trained model \cite{radford2021learning} and map into the same feature space $D$ by an embedding layer. 
\rev{As a decoder-only transformer architecture, }
the embedded text tokens are concatenated with the noised visual tokens. The transformer $f_{\theta}(\hat{x}_t; c, t)$ operates on these inputs, denoising the visual tokens at each timestep $t$. The denoised tokens $\hat{x}_0 \in \mathbb{R}^{K \times D}$ are subsequently decoded and reassembled into the visual data $x_0 \in \mathbb{R}^{T \times 3 \times H \times W}$ via linear layers, and finally decoded as videos or images. 

\new{To preserve the generalization ability of the finetuned t2v model across diverse prompts}, we curate two carefully selected datasets: (1) a 360° video dataset capturing static indoor environments, and (2) a 360° image dataset featuring a wide range of real-world scenes. \new{We also incorporate a masked diffusion loss, which is key to ensuring artifact-free generation.}

\subsubsection{360° video dataset}
We start by collecting a 360° video dataset with approximately 1,700 panoramic videos. Each video is captured by a person holding a Ricoh Theta Z1 camera walking inside a nearly static indoor scene to ensure the quality of 3D scene reconstruction. Each video has around 50 seconds of capture, yielding roughly 1,500 frames.
To build a dataset for training, we segment the longer video sequences into smaller chunks, typically a 128-frame video clip (5.3/10.6 seconds at 24/12fps). We use LLaVA \cite{liu2023improvedllava} to generate detailed captions describing each scene.

\subsubsection{360° image dataset}
However, as the 360° video data is primarily captured indoors to ensure a static environment with no moving objects during the capture process, this may bias the fine-tuned model, limiting its ability to generalize effectively to outdoor environments. To mitigate this issue, we gather another dataset with around 200,000 panoramic images taken with cameras on a tripod, without the photographer in the scene. We caption each image with Blip-2~\cite{li2023blip} to describe the whole scene in detail. These stationary images, captured in-the-wild, represent a much broader variety of scenes, thereby enhancing the generalization capability of our video diffusion model. 

\subsubsection{Mixed Training with Masked Loss}
We combine the two described datasets to train our model, enabling the text-to-360° video generator to produce both 360° videos and images, with images treated as single-frame videos. While panoramic images depict a clean environment without the photographer present, the 360° video capture often includes the photographer visible in the scene. This causes the trained model to produce camera operators in its outputs (see Fig. \ref{fig:pano_video_gen_ablation_mask}), complicating subsequent reconstruction steps. Hence, when training our diffusion models on video data, we introduce a \textit{masked} diffusion loss, where the photographer is excluded from the loss computation. 

To achieve this, we preprocess all 360° videos by masking out the photographer using a pretrained image segmentation model \cite{SelectSuject} on each frame in a video. For a 128-frame chunk, we merge all segmentation masks to minimize segmentation errors and ensure a clean region without the person. Additionally, we conservatively mask out the bottom region of the equirectangular panoramic frames to eliminate visible camera equipment and the photographer's hand. Assuming the corresponding binary mask $M$ is resized into the latent space by nearest neighbor interpolation, the masked diffusion loss is computed as 
\begin{equation}
   \mathbb{E}_{x \sim p_{\text{data}}, t \sim U(0,1)} \left[ \left\lVert M \odot \left( \epsilon_t - f_{\theta}(\hat{x}_t; c, t) \right) \right\rVert^2 \right],
    \label{eq:masked_objective}
\end{equation}
where $\epsilon_t$ is the sampled noise at timestep $t$, and \(M\) has the same dimensions as \(\epsilon_t\), where \(M[i, j] = 1\) includes the region in the loss calculation, and \(M[i, j] = 0\) excludes it (indicating the visible person). 

Our video generator is finally trained on a mix of $1/3$ image data and $2/3$ video data, achieving an empirically reasonable balance. The video diffusion model generates a 128-frame video clip at 12 fps, equivalent to a 10.6-second 360° capture video at a native resolution of 352$\times$704 pixels. We further apply a pretrained GAN-based video upsampler \cite{wang2018esrgan} to super-resolve our generated videos 4 times, resulting a final 360° video of 1416x2832 resolution. 

\subsection{Reconstruction from ``Generated'' Captures} \label{sec:recon}
Given the generated 360° capture video with 128 panoramic frames, we reconstruct a 3D Gaussian scene. For each panoramic frame, we randomly perform 3 perspective crops with a field of view of 120° and resolution of 512$\times$512, resulting in 384 perspective images for pose estimation and reconstruction.
Similar to standard scene reconstruction, we calibrate the camera poses of these perspective images using COLMAP \cite{schoenberger2016sfm}. The high average matching ratio $94.6\%$ from COLMAP estimation demonstrates the strong view consistency of the generated videos from our text-to-360° video generator. We adopt Long-LRM, the state-of-the-art feed-forward 3DGS reconstructor \cite{ziwen2024llrm} to reconstruct the final 3D scene using the estimated camera poses and generated views. Using Long-LRM allows us to achieve fast reconstruction compared to an optimization-based approach. We randomly subsample 32 random views along the original camera path as the inputs since the Long-LRM has a maximum limit for input views. The calibrated camera poses from COLMAP are converted into Plücker rays as the inputs for long-LRM.

\subsection{Implementation Details}
We fine-tune a pretrained text-to-video diffusion model using mixed 360° image and video datasets. The fine-tuning is conducted on 32 NVIDIA A100 GPUs \new{with a mixed fps training schedule (12fps and 24fps)}.
The model was finetuned from a pretrained text-to-video checkpoint for 20,000 steps in around 200 hours. During inference, we generate videos at 12 fps to extend the traversal range while maintaining good temporal consistency comparable to 24 fps generation. 

The inference of a 10.6-second video takes around 10 minutes with an unoptimized PyTorch runtime \cite{paszke2019pytorch} using 50 diffusion steps, together with the video upsampling process. The entire reconstruction takes around 5 minutes, including pose estimation and Long-LRM reconstruction. Our full 3D generation pipeline requires approximately 15 minutes to generate a detailed GS scene, with potential for further optimization to improve inference speed and efficiency for different components.

\begin{figure*}[ht]
    \centering
    \includegraphics[width=\textwidth]{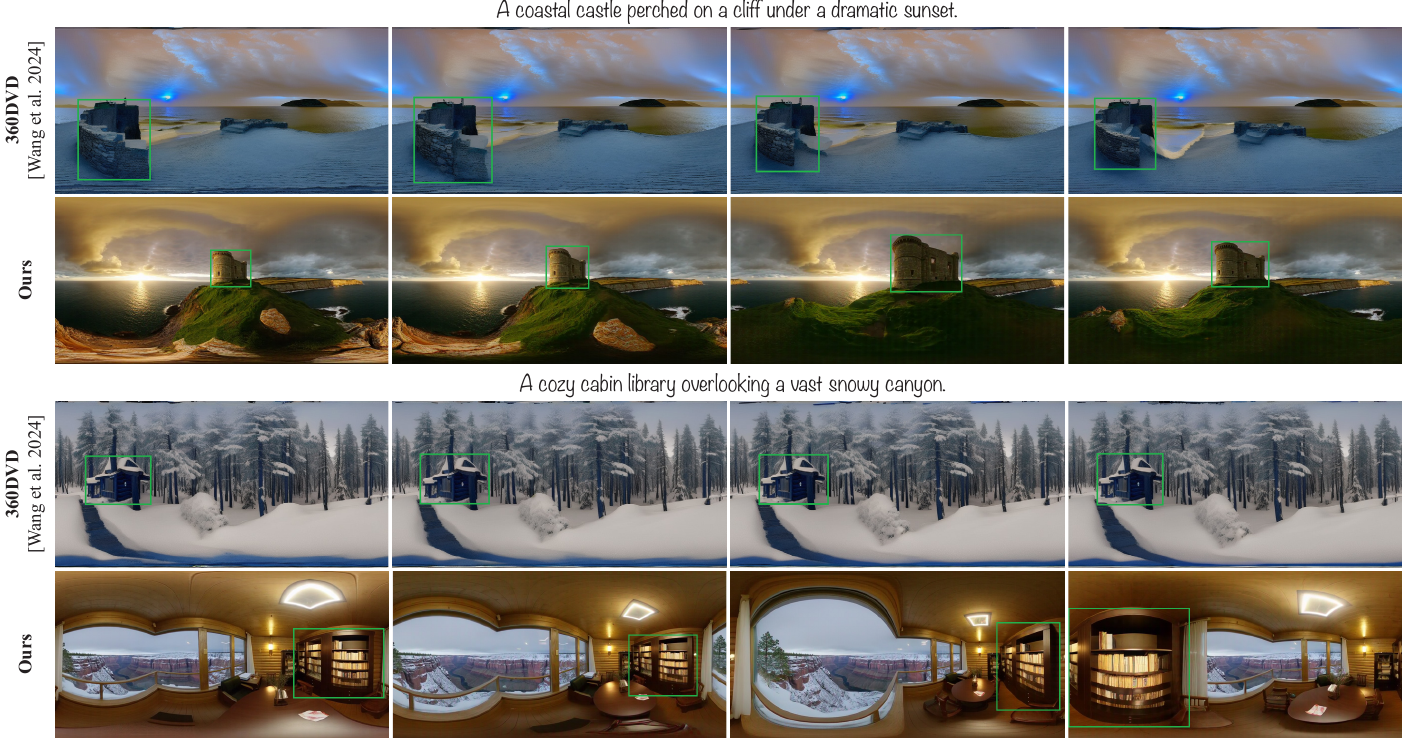}
    \caption{\textbf{Comparison of 360° video generation results.} We present a comparison of 360° panoramic video generation results against 360DVD \cite{wang2024360dvd}. Our generated videos exhibit significantly better visual quality and prompt alignment. We manually mark an anchor region with a green box in the 360° image, highlighting the substantially longer traversal range of our generated videos. \new{Stronger parallax is important for accurate 3D reconstruction.}
    }
    \label{fig:pano_video_gen_comparison}
\end{figure*}

\begin{figure}[t]
    \centering
    \includegraphics[width=0.45\textwidth]{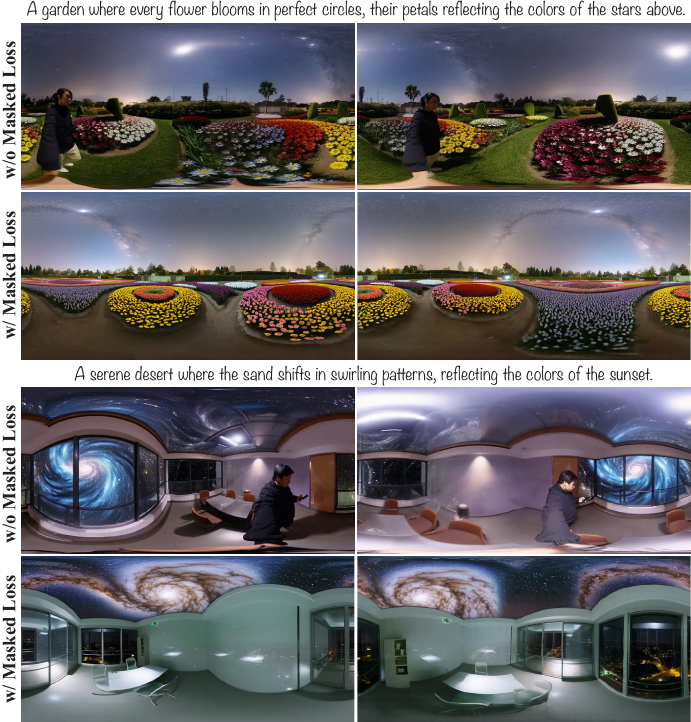}
    \vspace{-0.3cm}
    \caption{We conduct an ablation study to evaluate the impact of the masked diffusion loss. This loss function effectively guides the model to generate only the pixels of the scene, eliminating the moving photographer present in the original video training data. As a result, it enables the generation of a complete 360° scene capture for reconstruction.}
    \label{fig:pano_video_gen_ablation_mask}
\end{figure}

\section{Results}
\new{By generating 360° video capture, we demonstrate that our pipeline achieves state-of-the-art quality in 3D scene generation (Sec. \ref{sec:text2scene}), offering significantly improved view coverage, traversability and coherent global visual content. This is primarily due to a strong 360° video generator for scenes (Sec. \ref{sec:text2360})}. To quantitatively evaluate our method, we prompt ChatGPT \cite{OpenAI_ChatGPT} to generate a list of 320 descriptions of 3D scenes. \new{Using this in-the-wild test set, we conduct systematic comparisons against existing methods, demonstrating superior video and 3D scene generation quality through both visual examples and quantitative metrics. More visual examples can be found in the supplementary material.} 

Additionally, we perform ablation studies in Sec. \ref{sec:ablation} to evaluate our design choices, including mixed data training and masked loss, to ensure the generation of high-quality videos suitable for artifact-free 3D scene reconstruction.

\begin{table}[h]
    \footnotesize
    \centering
    \setlength{\tabcolsep}{4pt}   
    \caption{We quantitatively evaluate our method against the state-of-the-art (SOTA) 360° video generation model \cite{wang2024360dvd}. \new{We report CLIP distance (CLIP) for prompt alignment quality, Q-Align score for image quality, VBench for aesthetic/imaging quality, and COLMAP Failure Rate (COLMAP-FR) along with Matching Rate (COLMAP-MR) for multi-view consistency. The metrics are computed on the generated 360° frames across 320 test text prompts. Quantitative metrics across multiple dimensions demonstrate that our generated 360° videos achieve high visual quality and can be reliably used for 3D scene representation and reconstruction.}}
    \begin{tabular}{lccccc}
        \toprule
                        & CLIP$\downarrow$      & Q-Align$\uparrow$ & VBench$\uparrow$              & COLMAP-FR$\downarrow$     & COLMAP-MR$\uparrow$ \\
        \midrule
        360DVD          & 0.7643                & 0.6704            & 0.538/0.661                   &  89.7\%                   & 41.8\% \\
        \textbf{Ours}   & \textbf{0.7387}       & \textbf{0.8334}   & \textbf{0.615}/\textbf{0.717} & \textbf{0\%}              & \textbf{94.6\%} \\
        \bottomrule
    \end{tabular}
    \label{tab:comparison_video}
\end{table}

\begin{table}[h]
    \footnotesize
    \centering
    \setlength{\tabcolsep}{5.5pt}   
    \caption{\new{We quantitatively evaluate our method against several text-to-scene models, including \cite{chung2023luciddreamer}, \cite{zhou2025dreamscene360}, and \cite{yang2024layerpano3d}. To assess prompt adherence and visual quality, we compute CLIP distance and Q-Align metrics on the generated 360° images or frames across 320 test prompts randomly generated by ChatGPT. Note that LucidDreamer is an image-to-scene generation model which requires both text and image as inputs, and we generate the initial image using ChatGPT; therefore CLIP and Q-Align score evaluating the panorama synthesis quality is not applicable for LucidDreamer. Our method outperforms all baselines across these evaluations. Furthermore, we conduct a 4AFC user study to assess three key aspects of scene quality: Realism/Completeness/Traversal Consistency and report the selection percentages.}}
    \begin{tabular}{lccccccc}
        \toprule
                          & CLIP$\downarrow$     & Q-Align$\uparrow$  &  Real. $\uparrow$ & Complete. $\uparrow$ & Consist. $\uparrow$ \\
        \midrule
        LucidDreamer      & N/A                 & N/A                 & 21.0\% & 10.0\% & 10.0\%             \\
        DreamScene360     & 0.8116              & 0.8074              & 7.0\% & 6.0\% & 6.0\%             \\
        LayerPano3D       & 0.7977              & 0.8312              & 34.0\% & 25.0\% & 29.0\%             \\
        \textbf{Ours}     & \textbf{0.7387}     & \textbf{0.8334}     & \textbf{38.0\%} & \textbf{59.0\%} & \textbf{55.0\%}             \\
        \bottomrule
    \end{tabular}
    \label{tab:comparison_scene}
\end{table}

\subsection{Text-to-360°-Video Generation}\label{sec:text2360}
\new{A 360° video generator that accurately reproduces the distribution of real-world video captures is key to making it a sufficient proxy for 3D scene representation. } Our generation model can generate spatial-temporal consistent 360° panoramic videos, exceeding the quality of the state-of-the-art panoramic video generation models. We compare our 360° video generation results with 360DVD \cite{wang2024360dvd} in \figref{fig:pano_video_gen_comparison}. We showcase four evenly sampled frames comparing 360DVD and our 360° panorama video generation model. Our model generates videos that closely match the input text prompt and produce more realistic scene details compared to 360DVD. \new{In Table~\ref{tab:comparison_video}, we evaluate generation quality across multiple axes, where our generator consistently outperforms previous SOTA in common image/video generation quality metrics e.g., CLIP \cite{radford2021learning}, Q-align \cite{wu2023qalign}, VBench \cite{huang2023vbench}}.

Additionally, videos generated by our model cover a broader scene span compared to 360DVD, which exhibits only slight rotations and translations of the viewpoint as indicated in the marked green box (Fig. \ref{fig:pano_video_gen_comparison}). More importantly, since the primary goal of 360° panoramic video synthesis is to generate a coherent 3D scene, the quality of 360DVD's results prevents it from effectively accomplishing this task. Its outputs lack structural consistency across views, exhibiting continuous morphing and noticeable changes, which undermine the perception of a stable 3D environment.  

To validate this statement, we run COLMAP on their generated panoramas using a similar cropping and pose estimation procedure as described in Sec. \ref{sec:recon}. As shown in Table \ref{tab:comparison_video}, we find that the failure rate of COLMAP reaches $89.7\%$ (COLMAP-FR), and even in cases where COLMAP succeeds, the match rate (COLMAP-MR) remains very low, with only $41.8\%$ images being successfully matched, making their results unsuitable for 3D reconstruction. In contrast, our model preserves view-consistent local structures across frames. We do not encounter any failure cases when running COLMAP, and the average image match rate is as high as $94.6\%$ across our test prompts. This is the key reason our 360° videos can be a reasonable 3D representation for reconstruction.

\subsection{Text-to-Scene Generation} \label{sec:text2scene}
\begin{figure*}[ht]
    \centering
    \includegraphics[width=\textwidth]{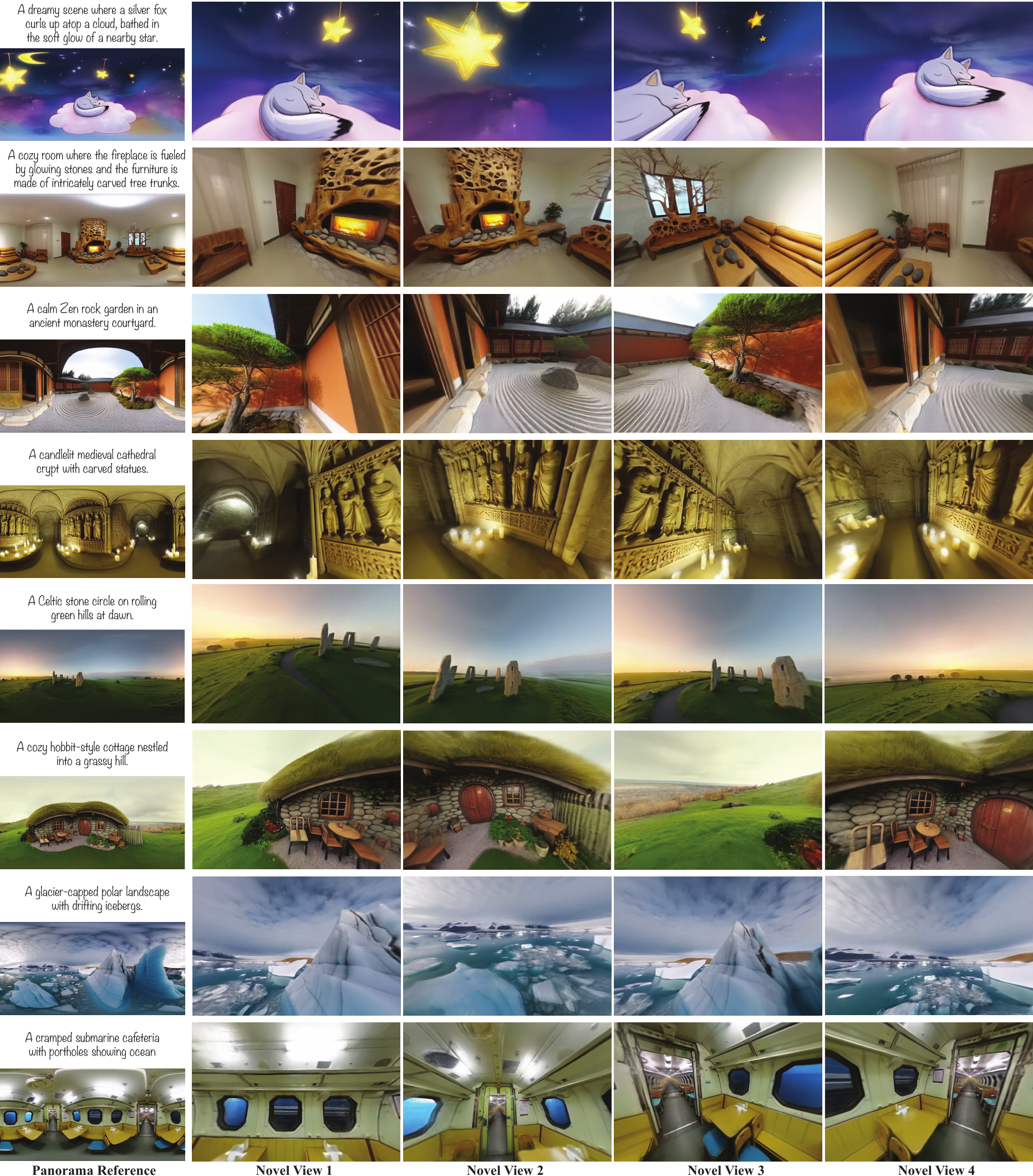}
    \vspace{-0.3cm}
    \caption{\textbf{Scene generation results.} We present our scene generation results by showing (1) the first 360° image from our generated panoramic video as reference, and (2) novel video renderings (Novel View 1-4) from different spatial locations and viewpoints, demonstrating that our approach effectively synthesizes a traversable, 360° viewable scene. More rendered videos can be found in the supplementary documents.}
    \label{fig:scene_gen_results}
\end{figure*}

As shown in Fig. \ref{fig:scene_gen_results}, by chaining our 360° video generator with the 3D reconstruction process, we enable the creation of a diverse set of 3D scenes, allowing users to navigate the environment from novel viewpoints and spatial locations. More results can be found in the supplementary material. 

In Fig. \ref{fig:scene_gen_comparison}, we compare our results to the previous state-of-the-art text-to-scene method that can also create 3D scenes. \new{LucidDreamer \cite{chung2023luciddreamer}, an outpainting-based scene generation method, synthesizes single-perspective views using an image diffusion model, but suffers from global geometric distortions and lacks support for full-view navigational experiences. Other methods that construct 3D scenes from panoramic images—such as DreamScene360 \cite{zhou2025dreamscene360} and LayerPano3D \cite{yang2024layerpano3d}—provide 360° view coverage, but do not enable truly navigable scenes, as significant camera movement often breaks scene coherence. For all other methods, when synthesizing novel views (Novel View 1–4 in Fig.~\ref{fig:scene_gen_comparison}) from positions different from the initial viewpoint, the rendered images exhibit significant artifacts, including blurriness, missing regions, and structural distortions.}  
Similarly, we report the CLIP distance and Q-Align metrics on the generated 360° images, demonstrating that our method generates a 3D environment that more accurately responds to user requests while achieving superior visual quality, even at the initial position.
\begin{figure}[t]
    \centering
    \includegraphics[width=0.45\textwidth]{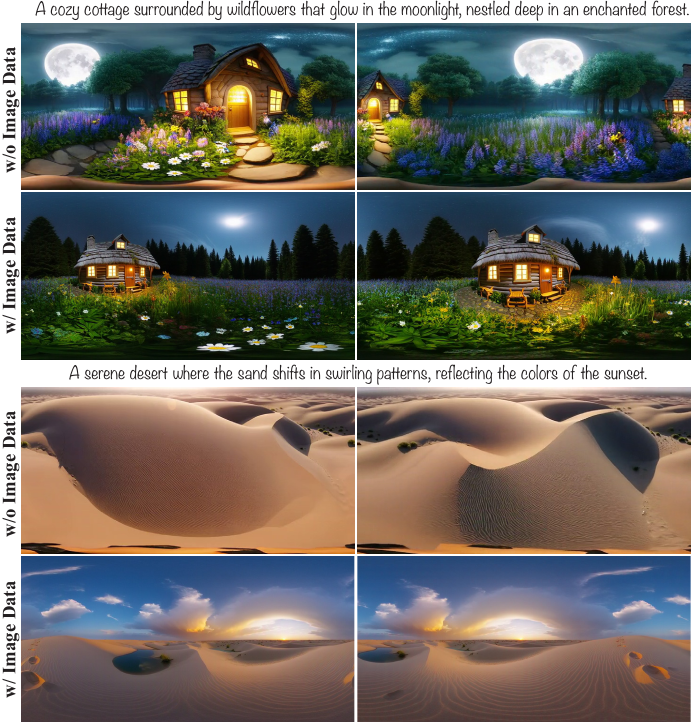}
    \vspace{-0.3cm}
    \caption{We verify that incorporating additional panorama image data helps the model generalize to diverse environments. Without the image data, the generated 360° videos not only exhibit degraded visual appearance but also display messy and incorrect content at the bottom of the video. This issue arises because these pixels lack supervision due to the masked loss.}
    \label{fig:pano_video_gen_ablation_image_data}
\end{figure}

\new{We argue that quality assessment is best conducted within the context of a real navigation experience. To this end, we perform a user study in which participants are presented with navigation videos generated by different methods. For each approach, we render a short video clip simulating camera movement through the scene, and ask users to select the most preferred result. Participants are instructed to evaluate the methods using a four-alternative forced-choice (4AFC) paradigm for the following dimensions: (1) Realism: Realistic Visual (2) Completeness: No missing regions (3) Traversal Consistency: Stable, distortion-free geometry during navigation. 
We collect 20 responses and report the selection percentages in Table~\ref{tab:comparison_scene}. Our method received the majority of votes across all evaluation criteria, with particularly strong margins in Completeness and Traversal Consistency.}

\subsection{Ablation Study} \label{sec:ablation}
Training a high-quality 360° video generator within our 3D generative pipeline is a key factor in the success of our method, relying on the use of mixed training data and the masked diffusion loss. We conduct ablation studies on these design choices to demonstrate their effectiveness.

\subsubsection{Mixed Training Data}
We conduct an experiment by excluding the additional image panorama data. As shown in Fig. \ref{fig:pano_video_gen_ablation_image_data}, incorporating image data significantly enhances the generalization capability of the video generator across a wide variety of scenes, and improves visual quality. Notably, due to the masked diffusion loss training, ground-truth pixels in the bottom regions of the panoramic videos are unavailable. The image data compensates for this limitation, enabling the model to intelligently auto-inpaint the bottom regions, resulting in a natural and realistic appearance.

\subsubsection{Masked Diffusion Loss}
We also train a model using the full diffusion loss applied to all pixels. The generated videos in this case reveal the photographer and the camera pod, which consistently appear in the 360° video training data (Fig. \ref{fig:pano_video_gen_ablation_mask}). This obstructs a significant portion of the generated capture videos, making comprehensive scene reconstruction challenging. By applying the masked diffusion loss and incorporating the aforementioned image panorama data, we guide the model to generate clean 360° videos that can be directly used for perspective cropping and reconstruction.

\rev{
\subsection{Discussions and Limitations}
The quality of our text-to-scene generation pipeline is constrained by both the length and resolution of the panoramic video generative model. On one hand, generating minute-long panoramic videos while maintaining visual consistency across extended sequences is essential for scaling to larger, more detailed 3D scenes. On the other hand, panoramic videos allocate more pixels to capture global 3D content and ensure spatial consistency, but this comes at the cost of fine-grained details compared to perspective videos at the same resolution. We address this limitation with an upsampler, though a more fundamental solution would be a high-resolution video generator.
}

\section{Conclusion}
\new{In conclusion, we demonstrate that 360° video capture is an effective and practical representation for scene-level 3D generation, as validated by \sysname, a generative pipeline capable of producing navigable 3D scenes from text prompts.} By leveraging 360° panoramic video generation as an intermediate representation, combined with a fast 3D reconstruction module, the pipeline enables the synthesis of immersive environments with comprehensive view coverage and high visual fidelity. This scalable approach effectively bridges the gap between text-to-3D generation and full-scene synthesis. \rev{
} 
\rev{We expect this novel idea highlights a promising research direction and has the potential to inspire future work in the field.}

\begin{figure*}[ht]
    \centering
    \includegraphics[width=\textwidth]{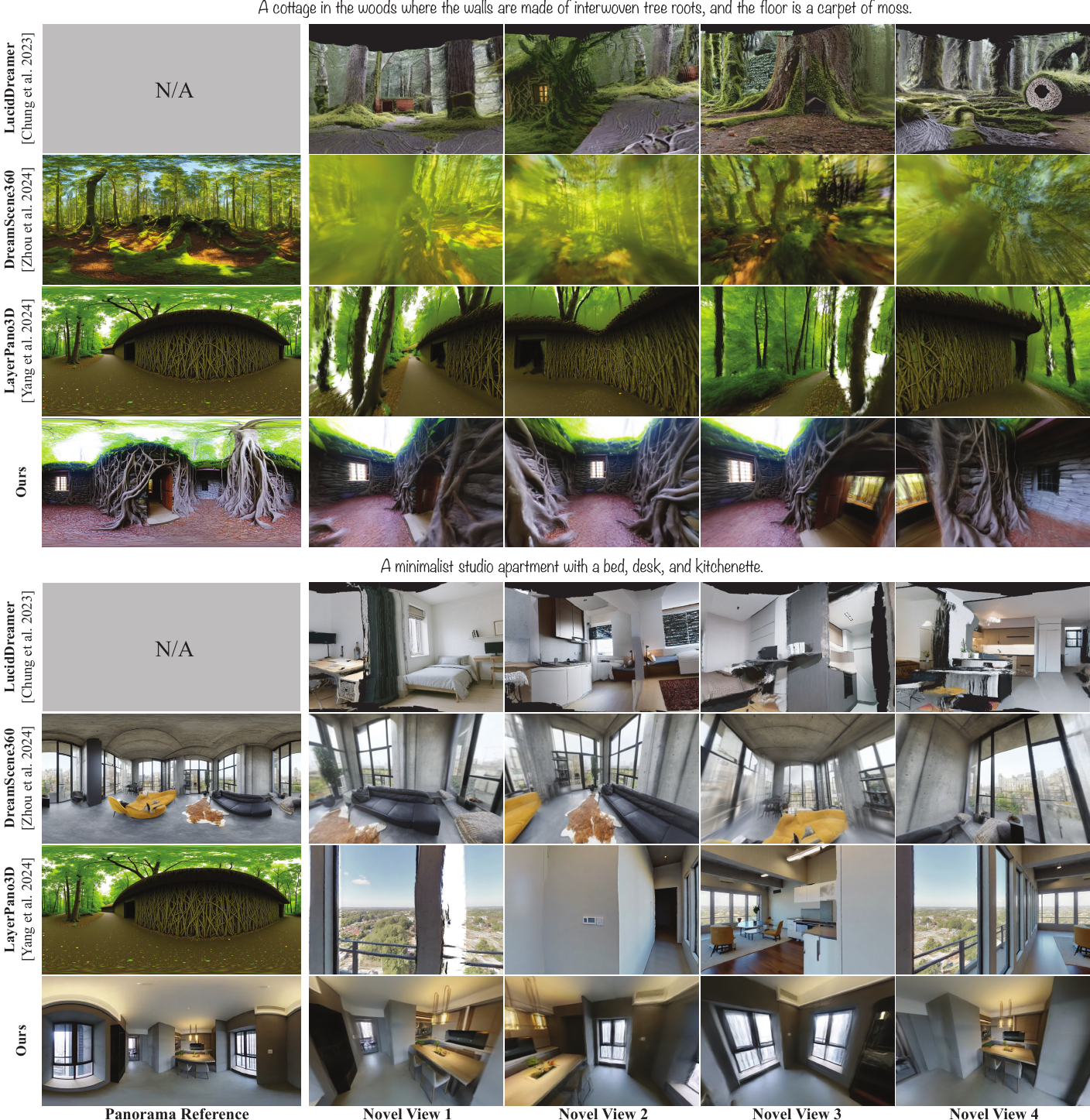}
    \caption{\textbf{Comparison of scene generation results.} We compare our method with LucidDreamer \cite{chung2023luciddreamer}, DreamScene360 \cite{zhou2025dreamscene360} and LayerPano3D \cite{yang2024layerpano3d} for text-based scene generation. 
    \new{LucidDreamer generates scenes progressively from a perspective images, synthesizing views one at a time. DreamScene360 and LayerPano3D both use a single panorama as input to an image diffusion model, shown in the first column for reference. For our method, we present the first frame of the generated 360° panoramic video as the reference image. While prior methods can produce visually reasonable results at the initial viewpoint, their range of motion is limited. When synthesizing novel viewpoints (Novel View 1–4) that deviate significantly from the starting position, their reconstructed Gaussian Splatting (GS) scenes exhibit severe artifacts, including blurriness, missing regions, and distorted geometries. As a result, global structural consistency cannot be held. In contrast, our method produces scenes that remain well-aligned with the input prompt and maintain high visual quality even at viewpoints far from the initial position.}}
    \label{fig:scene_gen_comparison}
\end{figure*}

\bibliographystyle{ACM-Reference-Format}
\bibliography{bibliography}


\begin{thebibliography}{71}


\ifx \showCODEN    \undefined \def \showCODEN     #1{\unskip}     \fi
\ifx \showISBNx    \undefined \def \showISBNx     #1{\unskip}     \fi
\ifx \showISBNxiii \undefined \def \showISBNxiii  #1{\unskip}     \fi
\ifx \showISSN     \undefined \def \showISSN      #1{\unskip}     \fi
\ifx \showLCCN     \undefined \def \showLCCN      #1{\unskip}     \fi
\ifx \shownote     \undefined \def \shownote      #1{#1}          \fi
\ifx \showarticletitle \undefined \def \showarticletitle #1{#1}   \fi
\ifx \showURL      \undefined \def \showURL       {\relax}        \fi
\providecommand\bibfield[2]{#2}
\providecommand\bibinfo[2]{#2}
\providecommand\natexlab[1]{#1}
\providecommand\showeprint[2][]{arXiv:#2}

\bibitem[Adobe(2025)]%
        {SelectSuject}
\bibfield{author}{\bibinfo{person}{Adobe}.} \bibinfo{year}{2025}\natexlab{}.
\newblock \bibinfo{title}{Select Subject}.
\newblock


\bibitem[Barron et~al\mbox{.}(2022)]%
        {barron2022mipnerf360unboundedantialiased}
\bibfield{author}{\bibinfo{person}{Jonathan~T. Barron}, \bibinfo{person}{Ben Mildenhall}, \bibinfo{person}{Dor Verbin}, \bibinfo{person}{Pratul~P. Srinivasan}, {and} \bibinfo{person}{Peter Hedman}.} \bibinfo{year}{2022}\natexlab{}.
\newblock \bibinfo{title}{Mip-NeRF 360: Unbounded Anti-Aliased Neural Radiance Fields}.
\newblock
\showeprint[arxiv]{2111.12077}~[cs.CV]
\urldef\tempurl%
\url{https://arxiv.org/abs/2111.12077}
\showURL{%
\tempurl}


\bibitem[Barron et~al\mbox{.}(2023)]%
        {barron2023zipnerfantialiasedgridbasedneural}
\bibfield{author}{\bibinfo{person}{Jonathan~T. Barron}, \bibinfo{person}{Ben Mildenhall}, \bibinfo{person}{Dor Verbin}, \bibinfo{person}{Pratul~P. Srinivasan}, {and} \bibinfo{person}{Peter Hedman}.} \bibinfo{year}{2023}\natexlab{}.
\newblock \bibinfo{title}{Zip-NeRF: Anti-Aliased Grid-Based Neural Radiance Fields}.
\newblock
\showeprint[arxiv]{2304.06706}~[cs.CV]
\urldef\tempurl%
\url{https://arxiv.org/abs/2304.06706}
\showURL{%
\tempurl}


\bibitem[Blattmann et~al\mbox{.}(2023a)]%
        {blattmann2023stablevideodiffusionscaling}
\bibfield{author}{\bibinfo{person}{Andreas Blattmann}, \bibinfo{person}{Tim Dockhorn}, \bibinfo{person}{Sumith Kulal}, \bibinfo{person}{Daniel Mendelevitch}, \bibinfo{person}{Maciej Kilian}, \bibinfo{person}{Dominik Lorenz}, \bibinfo{person}{Yam Levi}, \bibinfo{person}{Zion English}, \bibinfo{person}{Vikram Voleti}, \bibinfo{person}{Adam Letts}, \bibinfo{person}{Varun Jampani}, {and} \bibinfo{person}{Robin Rombach}.} \bibinfo{year}{2023}\natexlab{a}.
\newblock \bibinfo{title}{Stable Video Diffusion: Scaling Latent Video Diffusion Models to Large Datasets}.
\newblock
\showeprint[arxiv]{2311.15127}~[cs.CV]
\urldef\tempurl%
\url{https://arxiv.org/abs/2311.15127}
\showURL{%
\tempurl}


\bibitem[Blattmann et~al\mbox{.}(2023b)]%
        {blattmann2023align}
\bibfield{author}{\bibinfo{person}{Andreas Blattmann}, \bibinfo{person}{Robin Rombach}, \bibinfo{person}{Huan Ling}, \bibinfo{person}{Tim Dockhorn}, \bibinfo{person}{Seung~Wook Kim}, \bibinfo{person}{Sanja Fidler}, {and} \bibinfo{person}{Karsten Kreis}.} \bibinfo{year}{2023}\natexlab{b}.
\newblock \showarticletitle{Align your latents: High-resolution video synthesis with latent diffusion models}. In \bibinfo{booktitle}{\emph{Proceedings of the IEEE/CVF Conference on Computer Vision and Pattern Recognition}}. \bibinfo{pages}{22563--22575}.
\newblock


\bibitem[Brooks et~al\mbox{.}(2024)]%
        {videoworldsimulators2024}
\bibfield{author}{\bibinfo{person}{Tim Brooks}, \bibinfo{person}{Bill Peebles}, \bibinfo{person}{Connor Holmes}, \bibinfo{person}{Will DePue}, \bibinfo{person}{Yufei Guo}, \bibinfo{person}{Li Jing}, \bibinfo{person}{David Schnurr}, \bibinfo{person}{Joe Taylor}, \bibinfo{person}{Troy Luhman}, \bibinfo{person}{Eric Luhman}, \bibinfo{person}{Clarence Ng}, \bibinfo{person}{Ricky Wang}, {and} \bibinfo{person}{Aditya Ramesh}.} \bibinfo{year}{2024}\natexlab{}.
\newblock \showarticletitle{Video generation models as world simulators}.
\newblock  (\bibinfo{year}{2024}).
\newblock
\urldef\tempurl%
\url{https://openai.com/research/video-generation-models-as-world-simulators}
\showURL{%
\tempurl}


\bibitem[Charatan et~al\mbox{.}(2024)]%
        {charatan2024pixelsplat3dgaussiansplats}
\bibfield{author}{\bibinfo{person}{David Charatan}, \bibinfo{person}{Sizhe Li}, \bibinfo{person}{Andrea Tagliasacchi}, {and} \bibinfo{person}{Vincent Sitzmann}.} \bibinfo{year}{2024}\natexlab{}.
\newblock \bibinfo{title}{pixelSplat: 3D Gaussian Splats from Image Pairs for Scalable Generalizable 3D Reconstruction}.
\newblock
\showeprint[arxiv]{2312.12337}~[cs.CV]
\urldef\tempurl%
\url{https://arxiv.org/abs/2312.12337}
\showURL{%
\tempurl}


\bibitem[Chen et~al\mbox{.}(2022)]%
        {chen2022text2light}
\bibfield{author}{\bibinfo{person}{Zhaoxi Chen}, \bibinfo{person}{Guangcong Wang}, {and} \bibinfo{person}{Ziwei Liu}.} \bibinfo{year}{2022}\natexlab{}.
\newblock \showarticletitle{Text2Light: Zero-Shot Text-Driven HDR Panorama Generation}.
\newblock \bibinfo{journal}{\emph{ACM Transactions on Graphics (TOG)}} \bibinfo{volume}{41}, \bibinfo{number}{6}, Article \bibinfo{articleno}{195} (\bibinfo{year}{2022}), \bibinfo{numpages}{16}~pages.
\newblock


\bibitem[Chen et~al\mbox{.}(2024)]%
        {chen2024v3dvideodiffusionmodels}
\bibfield{author}{\bibinfo{person}{Zilong Chen}, \bibinfo{person}{Yikai Wang}, \bibinfo{person}{Feng Wang}, \bibinfo{person}{Zhengyi Wang}, {and} \bibinfo{person}{Huaping Liu}.} \bibinfo{year}{2024}\natexlab{}.
\newblock \bibinfo{title}{V3D: Video Diffusion Models are Effective 3D Generators}.
\newblock
\showeprint[arxiv]{2403.06738}~[cs.CV]
\urldef\tempurl%
\url{https://arxiv.org/abs/2403.06738}
\showURL{%
\tempurl}


\bibitem[Chung et~al\mbox{.}(2023)]%
        {chung2023luciddreamer}
\bibfield{author}{\bibinfo{person}{Jaeyoung Chung}, \bibinfo{person}{Suyoung Lee}, \bibinfo{person}{Hyeongjin Nam}, \bibinfo{person}{Jaerin Lee}, {and} \bibinfo{person}{Kyoung~Mu Lee}.} \bibinfo{year}{2023}\natexlab{}.
\newblock \showarticletitle{LucidDreamer: Domain-free Generation of 3D Gaussian Splatting Scenes}.
\newblock \bibinfo{journal}{\emph{arXiv preprint arXiv:2311.13384}} (\bibinfo{year}{2023}).
\newblock


\bibitem[Dastjerdi et~al\mbox{.}(2023)]%
        {dastjerdi2023everlight}
\bibfield{author}{\bibinfo{person}{Mohammad Reza~Karimi Dastjerdi}, \bibinfo{person}{Jonathan Eisenmann}, \bibinfo{person}{Yannick Hold-Geoffroy}, {and} \bibinfo{person}{Jean-Fran\c{c}ois Lalonde}.} \bibinfo{year}{2023}\natexlab{}.
\newblock \showarticletitle{EverLight: Indoor-Outdoor Editable {HDR} Lighting Estimation}. In \bibinfo{booktitle}{\emph{Proceedings of the IEEE/CVF International Conference on Computer Vision (ICCV)}}. \bibinfo{pages}{7420--7429}.
\newblock


\bibitem[Debevec and Malik(2023)]%
        {debevec2023recovering}
\bibfield{author}{\bibinfo{person}{Paul~E Debevec} {and} \bibinfo{person}{Jitendra Malik}.} \bibinfo{year}{2023}\natexlab{}.
\newblock \showarticletitle{Recovering high dynamic range radiance maps from photographs}.
\newblock In \bibinfo{booktitle}{\emph{Seminal Graphics Papers: Pushing the Boundaries, Volume 2}}. \bibinfo{pages}{643--652}.
\newblock


\bibitem[Esser et~al\mbox{.}(2024)]%
        {esser2024scaling}
\bibfield{author}{\bibinfo{person}{Patrick Esser}, \bibinfo{person}{Sumith Kulal}, \bibinfo{person}{Andreas Blattmann}, \bibinfo{person}{Rahim Entezari}, \bibinfo{person}{Jonas M{\"u}ller}, \bibinfo{person}{Harry Saini}, \bibinfo{person}{Yam Levi}, \bibinfo{person}{Dominik Lorenz}, \bibinfo{person}{Axel Sauer}, \bibinfo{person}{Frederic Boesel}, {et~al\mbox{.}}} \bibinfo{year}{2024}\natexlab{}.
\newblock \showarticletitle{Scaling rectified flow transformers for high-resolution image synthesis}. In \bibinfo{booktitle}{\emph{Forty-first International Conference on Machine Learning}}.
\newblock


\bibitem[Gao et~al\mbox{.}(2024)]%
        {MeshGaussian2024}
\bibfield{author}{\bibinfo{person}{Lin Gao}, \bibinfo{person}{Jie Yang}, \bibinfo{person}{Botao Zhang}, \bibinfo{person}{Jiamu Sun}, \bibinfo{person}{Yujie Yuan}, \bibinfo{person}{Hongbo Fu}, {and} \bibinfo{person}{Yu-Kun Lai}.} \bibinfo{year}{2024}\natexlab{}.
\newblock \showarticletitle{Real-time Large-scale Deformation of Gaussian Splatting}.
\newblock \bibinfo{journal}{\emph{ACM Transactions on Graphics (SIGGRAPH Asia 2024)}} (\bibinfo{year}{2024}).
\newblock


\bibitem[Han et~al\mbox{.}(2024)]%
        {han2024vfusion3dlearningscalable3d}
\bibfield{author}{\bibinfo{person}{Junlin Han}, \bibinfo{person}{Filippos Kokkinos}, {and} \bibinfo{person}{Philip Torr}.} \bibinfo{year}{2024}\natexlab{}.
\newblock \bibinfo{title}{VFusion3D: Learning Scalable 3D Generative Models from Video Diffusion Models}.
\newblock
\showeprint[arxiv]{2403.12034}~[cs.CV]
\urldef\tempurl%
\url{https://arxiv.org/abs/2403.12034}
\showURL{%
\tempurl}


\bibitem[H{\"o}llein et~al\mbox{.}(2023)]%
        {hollein2023text2room}
\bibfield{author}{\bibinfo{person}{Lukas H{\"o}llein}, \bibinfo{person}{Ang Cao}, \bibinfo{person}{Andrew Owens}, \bibinfo{person}{Justin Johnson}, {and} \bibinfo{person}{Matthias Nie{\ss}ner}.} \bibinfo{year}{2023}\natexlab{}.
\newblock \showarticletitle{Text2room: Extracting textured 3d meshes from 2d text-to-image models}. In \bibinfo{booktitle}{\emph{Proceedings of the IEEE/CVF International Conference on Computer Vision}}. \bibinfo{pages}{7909--7920}.
\newblock


\bibitem[Hu et~al\mbox{.}(2021)]%
        {hu2021loralowrankadaptationlarge}
\bibfield{author}{\bibinfo{person}{Edward~J. Hu}, \bibinfo{person}{Yelong Shen}, \bibinfo{person}{Phillip Wallis}, \bibinfo{person}{Zeyuan Allen-Zhu}, \bibinfo{person}{Yuanzhi Li}, \bibinfo{person}{Shean Wang}, \bibinfo{person}{Lu Wang}, {and} \bibinfo{person}{Weizhu Chen}.} \bibinfo{year}{2021}\natexlab{}.
\newblock \bibinfo{title}{LoRA: Low-Rank Adaptation of Large Language Models}.
\newblock
\showeprint[arxiv]{2106.09685}~[cs.CL]
\urldef\tempurl%
\url{https://arxiv.org/abs/2106.09685}
\showURL{%
\tempurl}


\bibitem[Hu et~al\mbox{.}(2024)]%
        {Hu_2024_CVPR}
\bibfield{author}{\bibinfo{person}{Hanzhe Hu}, \bibinfo{person}{Zhizhuo Zhou}, \bibinfo{person}{Varun Jampani}, {and} \bibinfo{person}{Shubham Tulsiani}.} \bibinfo{year}{2024}\natexlab{}.
\newblock \showarticletitle{MVD-Fusion: Single-view 3D via Depth-consistent Multi-view Generation}. In \bibinfo{booktitle}{\emph{Proceedings of the IEEE/CVF Conference on Computer Vision and Pattern Recognition (CVPR)}}. \bibinfo{pages}{9698--9707}.
\newblock


\bibitem[Huang et~al\mbox{.}(2024)]%
        {huang2023vbench}
\bibfield{author}{\bibinfo{person}{Ziqi Huang}, \bibinfo{person}{Yinan He}, \bibinfo{person}{Jiashuo Yu}, \bibinfo{person}{Fan Zhang}, \bibinfo{person}{Chenyang Si}, \bibinfo{person}{Yuming Jiang}, \bibinfo{person}{Yuanhan Zhang}, \bibinfo{person}{Tianxing Wu}, \bibinfo{person}{Qingyang Jin}, \bibinfo{person}{Nattapol Chanpaisit}, \bibinfo{person}{Yaohui Wang}, \bibinfo{person}{Xinyuan Chen}, \bibinfo{person}{Limin Wang}, \bibinfo{person}{Dahua Lin}, \bibinfo{person}{Yu Qiao}, {and} \bibinfo{person}{Ziwei Liu}.} \bibinfo{year}{2024}\natexlab{}.
\newblock \showarticletitle{{VBench}: Comprehensive Benchmark Suite for Video Generative Models}. In \bibinfo{booktitle}{\emph{Proceedings of the IEEE/CVF Conference on Computer Vision and Pattern Recognition}}.
\newblock


\bibitem[Jiang et~al\mbox{.}(2024)]%
        {COGS2024}
\bibfield{author}{\bibinfo{person}{Kaiwen Jiang}, \bibinfo{person}{Yang Fu}, \bibinfo{person}{Mukund Varma~T}, \bibinfo{person}{Yash Belhe}, \bibinfo{person}{Xiaolong Wang}, \bibinfo{person}{Hao Su}, {and} \bibinfo{person}{Ravi Ramamoorthi}.} \bibinfo{year}{2024}\natexlab{}.
\newblock \showarticletitle{A Construct-Optimize Approach to Sparse View Synthesis without Camera Pose}.
\newblock \bibinfo{journal}{\emph{SIGGRAPH}} (\bibinfo{year}{2024}).
\newblock


\bibitem[Kazhdan et~al\mbox{.}(2006)]%
        {kazhdan2006poisson}
\bibfield{author}{\bibinfo{person}{Michael Kazhdan}, \bibinfo{person}{Matthew Bolitho}, {and} \bibinfo{person}{Hugues Hoppe}.} \bibinfo{year}{2006}\natexlab{}.
\newblock \showarticletitle{Poisson surface reconstruction}. In \bibinfo{booktitle}{\emph{Proceedings of the fourth Eurographics symposium on Geometry processing}}, Vol.~\bibinfo{volume}{7}.
\newblock


\bibitem[Kerbl et~al\mbox{.}(2023)]%
        {kerbl3Dgaussians}
\bibfield{author}{\bibinfo{person}{Bernhard Kerbl}, \bibinfo{person}{Georgios Kopanas}, \bibinfo{person}{Thomas Leimk{\"u}hler}, {and} \bibinfo{person}{George Drettakis}.} \bibinfo{year}{2023}\natexlab{}.
\newblock \showarticletitle{3D Gaussian Splatting for Real-Time Radiance Field Rendering}.
\newblock \bibinfo{journal}{\emph{ACM Transactions on Graphics}} \bibinfo{volume}{42}, \bibinfo{number}{4} (\bibinfo{date}{July} \bibinfo{year}{2023}).
\newblock
\urldef\tempurl%
\url{https://repo-sam.inria.fr/fungraph/3d-gaussian-splatting/}
\showURL{%
\tempurl}


\bibitem[Li et~al\mbox{.}(2023a)]%
        {li2023blip}
\bibfield{author}{\bibinfo{person}{Junnan Li}, \bibinfo{person}{Dongxu Li}, \bibinfo{person}{Silvio Savarese}, {and} \bibinfo{person}{Steven Hoi}.} \bibinfo{year}{2023}\natexlab{a}.
\newblock \showarticletitle{Blip-2: Bootstrapping language-image pre-training with frozen image encoders and large language models}. In \bibinfo{booktitle}{\emph{International conference on machine learning}}. PMLR, \bibinfo{pages}{19730--19742}.
\newblock


\bibitem[Li et~al\mbox{.}(2023b)]%
        {li2023instant3dfasttextto3dsparseview}
\bibfield{author}{\bibinfo{person}{Jiahao Li}, \bibinfo{person}{Hao Tan}, \bibinfo{person}{Kai Zhang}, \bibinfo{person}{Zexiang Xu}, \bibinfo{person}{Fujun Luan}, \bibinfo{person}{Yinghao Xu}, \bibinfo{person}{Yicong Hong}, \bibinfo{person}{Kalyan Sunkavalli}, \bibinfo{person}{Greg Shakhnarovich}, {and} \bibinfo{person}{Sai Bi}.} \bibinfo{year}{2023}\natexlab{b}.
\newblock \bibinfo{title}{Instant3D: Fast Text-to-3D with Sparse-View Generation and Large Reconstruction Model}.
\newblock
\showeprint[arxiv]{2311.06214}~[cs.CV]
\urldef\tempurl%
\url{https://arxiv.org/abs/2311.06214}
\showURL{%
\tempurl}


\bibitem[Li et~al\mbox{.}(2024b)]%
        {li20244k4dgenpanoramic4dgeneration}
\bibfield{author}{\bibinfo{person}{Renjie Li}, \bibinfo{person}{Panwang Pan}, \bibinfo{person}{Bangbang Yang}, \bibinfo{person}{Dejia Xu}, \bibinfo{person}{Shijie Zhou}, \bibinfo{person}{Xuanyang Zhang}, \bibinfo{person}{Zeming Li}, \bibinfo{person}{Achuta Kadambi}, \bibinfo{person}{Zhangyang Wang}, \bibinfo{person}{Zhengzhong Tu}, {and} \bibinfo{person}{Zhiwen Fan}.} \bibinfo{year}{2024}\natexlab{b}.
\newblock \bibinfo{title}{4K4DGen: Panoramic 4D Generation at 4K Resolution}.
\newblock
\showeprint[arxiv]{2406.13527}~[cs.CV]
\urldef\tempurl%
\url{https://arxiv.org/abs/2406.13527}
\showURL{%
\tempurl}


\bibitem[Li et~al\mbox{.}(2024a)]%
        {li2024director3d}
\bibfield{author}{\bibinfo{person}{Xinyang Li}, \bibinfo{person}{Zhangyu Lai}, \bibinfo{person}{Linning Xu}, \bibinfo{person}{Yansong Qu}, \bibinfo{person}{Liujuan Cao}, \bibinfo{person}{Shengchuan Zhang}, \bibinfo{person}{Bo Dai}, {and} \bibinfo{person}{Rongrong Ji}.} \bibinfo{year}{2024}\natexlab{a}.
\newblock \showarticletitle{Director3D: Real-world Camera Trajectory and 3D Scene Generation from Text}.
\newblock \bibinfo{journal}{\emph{arXiv:2406.17601}} (\bibinfo{year}{2024}).
\newblock


\bibitem[Liu et~al\mbox{.}(2023a)]%
        {liu2023improvedllava}
\bibfield{author}{\bibinfo{person}{Haotian Liu}, \bibinfo{person}{Chunyuan Li}, \bibinfo{person}{Yuheng Li}, {and} \bibinfo{person}{Yong~Jae Lee}.} \bibinfo{year}{2023}\natexlab{a}.
\newblock \bibinfo{title}{Improved Baselines with Visual Instruction Tuning}.
\newblock


\bibitem[Liu et~al\mbox{.}(2023b)]%
        {liu2023llava}
\bibfield{author}{\bibinfo{person}{Haotian Liu}, \bibinfo{person}{Chunyuan Li}, \bibinfo{person}{Qingyang Wu}, {and} \bibinfo{person}{Yong~Jae Lee}.} \bibinfo{year}{2023}\natexlab{b}.
\newblock \showarticletitle{Visual Instruction Tuning}. In \bibinfo{booktitle}{\emph{NeurIPS}}.
\newblock


\bibitem[Liu et~al\mbox{.}(2024b)]%
        {liu2023one2345}
\bibfield{author}{\bibinfo{person}{Minghua Liu}, \bibinfo{person}{Chao Xu}, \bibinfo{person}{Haian Jin}, \bibinfo{person}{Linghao Chen}, \bibinfo{person}{Mukund Varma~T}, \bibinfo{person}{Zexiang Xu}, {and} \bibinfo{person}{Hao Su}.} \bibinfo{year}{2024}\natexlab{b}.
\newblock \showarticletitle{One-2-3-45: Any single image to 3d mesh in 45 seconds without per-shape optimization}.
\newblock \bibinfo{journal}{\emph{Advances in Neural Information Processing Systems}}  \bibinfo{volume}{36} (\bibinfo{year}{2024}).
\newblock


\bibitem[Liu et~al\mbox{.}(2023c)]%
        {liu2023zero1to3zeroshotimage3d}
\bibfield{author}{\bibinfo{person}{Ruoshi Liu}, \bibinfo{person}{Rundi Wu}, \bibinfo{person}{Basile~Van Hoorick}, \bibinfo{person}{Pavel Tokmakov}, \bibinfo{person}{Sergey Zakharov}, {and} \bibinfo{person}{Carl Vondrick}.} \bibinfo{year}{2023}\natexlab{c}.
\newblock \bibinfo{title}{Zero-1-to-3: Zero-shot One Image to 3D Object}.
\newblock
\showeprint[arxiv]{2303.11328}~[cs.CV]
\urldef\tempurl%
\url{https://arxiv.org/abs/2303.11328}
\showURL{%
\tempurl}


\bibitem[Liu et~al\mbox{.}(2024a)]%
        {liu2024pyramiddiffusionfine3d}
\bibfield{author}{\bibinfo{person}{Yuheng Liu}, \bibinfo{person}{Xinke Li}, \bibinfo{person}{Xueting Li}, \bibinfo{person}{Lu Qi}, \bibinfo{person}{Chongshou Li}, {and} \bibinfo{person}{Ming-Hsuan Yang}.} \bibinfo{year}{2024}\natexlab{a}.
\newblock \showarticletitle{Pyramid Diffusion for Fine 3D Large Scene Generation}. In \bibinfo{booktitle}{\emph{European Conference on Computer Vision (ECCV)}}.
\newblock


\bibitem[Mildenhall et~al\mbox{.}(2020)]%
        {mildenhall2020nerf}
\bibfield{author}{\bibinfo{person}{Ben Mildenhall}, \bibinfo{person}{Pratul~P. Srinivasan}, \bibinfo{person}{Matthew Tancik}, \bibinfo{person}{Jonathan~T. Barron}, \bibinfo{person}{Ravi Ramamoorthi}, {and} \bibinfo{person}{Ren Ng}.} \bibinfo{year}{2020}\natexlab{}.
\newblock \showarticletitle{NeRF: Representing Scenes as Neural Radiance Fields for View Synthesis}. In \bibinfo{booktitle}{\emph{ECCV}}.
\newblock


\bibitem[M\"uller et~al\mbox{.}(2022)]%
        {mueller2022instant}
\bibfield{author}{\bibinfo{person}{Thomas M\"uller}, \bibinfo{person}{Alex Evans}, \bibinfo{person}{Christoph Schied}, {and} \bibinfo{person}{Alexander Keller}.} \bibinfo{year}{2022}\natexlab{}.
\newblock \showarticletitle{Instant Neural Graphics Primitives with a Multiresolution Hash Encoding}.
\newblock \bibinfo{journal}{\emph{ACM Trans. Graph.}} \bibinfo{volume}{41}, \bibinfo{number}{4}, Article \bibinfo{articleno}{102} (\bibinfo{date}{July} \bibinfo{year}{2022}), \bibinfo{numpages}{15}~pages.
\newblock
\href{https://doi.org/10.1145/3528223.3530127}{doi:\nolinkurl{10.1145/3528223.3530127}}


\bibitem[OpenAI(2023)]%
        {OpenAI_ChatGPT}
\bibfield{author}{\bibinfo{person}{OpenAI}.} \bibinfo{year}{2023}\natexlab{}.
\newblock \bibinfo{title}{ChatGPT: A Large Language Model}.
\newblock
\urldef\tempurl%
\url{https://chat.openai.com}
\showURL{%
\tempurl}
\newblock
\shownote{Accessed: 2025-03-07}.


\bibitem[Ouyang et~al\mbox{.}(2023)]%
        {ouyang2023text}
\bibfield{author}{\bibinfo{person}{Hao Ouyang}, \bibinfo{person}{Tiancheng Sun}, \bibinfo{person}{Stephen Lombardi}, {and} \bibinfo{person}{Kathryn Heal}.} \bibinfo{year}{2023}\natexlab{}.
\newblock \showarticletitle{Text2Immersion: Generative Immersive Scene with 3D Gaussians}.
\newblock \bibinfo{journal}{\emph{Arxiv}} (\bibinfo{year}{2023}).
\newblock


\bibitem[Parthasarathy et~al\mbox{.}(2024)]%
        {parthasarathy2024vid3dsynthesisdynamic3d}
\bibfield{author}{\bibinfo{person}{Rishab Parthasarathy}, \bibinfo{person}{Zachary Ankner}, {and} \bibinfo{person}{Aaron Gokaslan}.} \bibinfo{year}{2024}\natexlab{}.
\newblock \bibinfo{title}{Vid3D: Synthesis of Dynamic 3D Scenes using 2D Video Diffusion}.
\newblock
\showeprint[arxiv]{2406.11196}~[cs.CV]
\urldef\tempurl%
\url{https://arxiv.org/abs/2406.11196}
\showURL{%
\tempurl}


\bibitem[Paszke et~al\mbox{.}(2019)]%
        {paszke2019pytorch}
\bibfield{author}{\bibinfo{person}{Adam Paszke}, \bibinfo{person}{Sam Gross}, \bibinfo{person}{Francisco Massa}, \bibinfo{person}{Adam Lerer}, \bibinfo{person}{James Bradbury}, \bibinfo{person}{Gregory Chanan}, \bibinfo{person}{Trevor Killeen}, \bibinfo{person}{Zeming Lin}, \bibinfo{person}{Natalia Gimelshein}, \bibinfo{person}{Luca Antiga}, {et~al\mbox{.}}} \bibinfo{year}{2019}\natexlab{}.
\newblock \showarticletitle{Pytorch: An imperative style, high-performance deep learning library}.
\newblock \bibinfo{journal}{\emph{Advances in neural information processing systems}}  \bibinfo{volume}{32} (\bibinfo{year}{2019}).
\newblock


\bibitem[Peebles and Xie(2023)]%
        {peebles2023scalable}
\bibfield{author}{\bibinfo{person}{William Peebles} {and} \bibinfo{person}{Saining Xie}.} \bibinfo{year}{2023}\natexlab{}.
\newblock \showarticletitle{Scalable diffusion models with transformers}. In \bibinfo{booktitle}{\emph{Proceedings of the IEEE/CVF International Conference on Computer Vision}}. \bibinfo{pages}{4195--4205}.
\newblock


\bibitem[Polyak et~al\mbox{.}(2024)]%
        {polyak2024moviegencastmedia}
\bibfield{author}{\bibinfo{person}{Adam Polyak}, \bibinfo{person}{Amit Zohar}, \bibinfo{person}{Andrew Brown}, \bibinfo{person}{Andros Tjandra}, \bibinfo{person}{Animesh Sinha}, \bibinfo{person}{Ann Lee}, \bibinfo{person}{Apoorv Vyas}, \bibinfo{person}{Bowen Shi}, \bibinfo{person}{Chih-Yao Ma}, \bibinfo{person}{Ching-Yao Chuang}, \bibinfo{person}{David Yan}, \bibinfo{person}{Dhruv Choudhary}, \bibinfo{person}{Dingkang Wang}, \bibinfo{person}{Geet Sethi}, \bibinfo{person}{Guan Pang}, \bibinfo{person}{Haoyu Ma}, \bibinfo{person}{Ishan Misra}, \bibinfo{person}{Ji Hou}, \bibinfo{person}{Jialiang Wang}, \bibinfo{person}{Kiran Jagadeesh}, \bibinfo{person}{Kunpeng Li}, \bibinfo{person}{Luxin Zhang}, \bibinfo{person}{Mannat Singh}, \bibinfo{person}{Mary Williamson}, \bibinfo{person}{Matt Le}, \bibinfo{person}{Matthew Yu}, \bibinfo{person}{Mitesh~Kumar Singh}, \bibinfo{person}{Peizhao Zhang}, \bibinfo{person}{Peter Vajda}, \bibinfo{person}{Quentin Duval}, \bibinfo{person}{Rohit Girdhar}, \bibinfo{person}{Roshan
  Sumbaly}, \bibinfo{person}{Sai~Saketh Rambhatla}, \bibinfo{person}{Sam Tsai}, \bibinfo{person}{Samaneh Azadi}, \bibinfo{person}{Samyak Datta}, \bibinfo{person}{Sanyuan Chen}, \bibinfo{person}{Sean Bell}, \bibinfo{person}{Sharadh Ramaswamy}, \bibinfo{person}{Shelly Sheynin}, \bibinfo{person}{Siddharth Bhattacharya}, \bibinfo{person}{Simran Motwani}, \bibinfo{person}{Tao Xu}, \bibinfo{person}{Tianhe Li}, \bibinfo{person}{Tingbo Hou}, \bibinfo{person}{Wei-Ning Hsu}, \bibinfo{person}{Xi Yin}, \bibinfo{person}{Xiaoliang Dai}, \bibinfo{person}{Yaniv Taigman}, \bibinfo{person}{Yaqiao Luo}, \bibinfo{person}{Yen-Cheng Liu}, \bibinfo{person}{Yi-Chiao Wu}, \bibinfo{person}{Yue Zhao}, \bibinfo{person}{Yuval Kirstain}, \bibinfo{person}{Zecheng He}, \bibinfo{person}{Zijian He}, \bibinfo{person}{Albert Pumarola}, \bibinfo{person}{Ali Thabet}, \bibinfo{person}{Artsiom Sanakoyeu}, \bibinfo{person}{Arun Mallya}, \bibinfo{person}{Baishan Guo}, \bibinfo{person}{Boris Araya}, \bibinfo{person}{Breena Kerr},
  \bibinfo{person}{Carleigh Wood}, \bibinfo{person}{Ce Liu}, \bibinfo{person}{Cen Peng}, \bibinfo{person}{Dimitry Vengertsev}, \bibinfo{person}{Edgar Schonfeld}, \bibinfo{person}{Elliot Blanchard}, \bibinfo{person}{Felix Juefei-Xu}, \bibinfo{person}{Fraylie Nord}, \bibinfo{person}{Jeff Liang}, \bibinfo{person}{John Hoffman}, \bibinfo{person}{Jonas Kohler}, \bibinfo{person}{Kaolin Fire}, \bibinfo{person}{Karthik Sivakumar}, \bibinfo{person}{Lawrence Chen}, \bibinfo{person}{Licheng Yu}, \bibinfo{person}{Luya Gao}, \bibinfo{person}{Markos Georgopoulos}, \bibinfo{person}{Rashel Moritz}, \bibinfo{person}{Sara~K. Sampson}, \bibinfo{person}{Shikai Li}, \bibinfo{person}{Simone Parmeggiani}, \bibinfo{person}{Steve Fine}, \bibinfo{person}{Tara Fowler}, \bibinfo{person}{Vladan Petrovic}, {and} \bibinfo{person}{Yuming Du}.} \bibinfo{year}{2024}\natexlab{}.
\newblock \bibinfo{title}{Movie Gen: A Cast of Media Foundation Models}.
\newblock
\showeprint[arxiv]{2410.13720}~[cs.CV]
\urldef\tempurl%
\url{https://arxiv.org/abs/2410.13720}
\showURL{%
\tempurl}


\bibitem[Qi et~al\mbox{.}(2017)]%
        {qi2017pointnet}
\bibfield{author}{\bibinfo{person}{Charles~R Qi}, \bibinfo{person}{Hao Su}, \bibinfo{person}{Kaichun Mo}, {and} \bibinfo{person}{Leonidas~J Guibas}.} \bibinfo{year}{2017}\natexlab{}.
\newblock \showarticletitle{Pointnet: Deep learning on point sets for 3d classification and segmentation}. In \bibinfo{booktitle}{\emph{Proceedings of the IEEE conference on computer vision and pattern recognition}}. \bibinfo{pages}{652--660}.
\newblock


\bibitem[Radford et~al\mbox{.}(2021)]%
        {radford2021learning}
\bibfield{author}{\bibinfo{person}{Alec Radford}, \bibinfo{person}{Jong~Wook Kim}, \bibinfo{person}{Chris Hallacy}, \bibinfo{person}{Aditya Ramesh}, \bibinfo{person}{Gabriel Goh}, \bibinfo{person}{Sandhini Agarwal}, \bibinfo{person}{Girish Sastry}, \bibinfo{person}{Amanda Askell}, \bibinfo{person}{Pamela Mishkin}, \bibinfo{person}{Jack Clark}, {et~al\mbox{.}}} \bibinfo{year}{2021}\natexlab{}.
\newblock \showarticletitle{Learning transferable visual models from natural language supervision}. In \bibinfo{booktitle}{\emph{International conference on machine learning}}. PMLR, \bibinfo{pages}{8748--8763}.
\newblock


\bibitem[Rombach et~al\mbox{.}(2022)]%
        {rombach2022highresolutionimagesynthesislatent}
\bibfield{author}{\bibinfo{person}{Robin Rombach}, \bibinfo{person}{Andreas Blattmann}, \bibinfo{person}{Dominik Lorenz}, \bibinfo{person}{Patrick Esser}, {and} \bibinfo{person}{Björn Ommer}.} \bibinfo{year}{2022}\natexlab{}.
\newblock \bibinfo{title}{High-Resolution Image Synthesis with Latent Diffusion Models}.
\newblock
\showeprint[arxiv]{2112.10752}~[cs.CV]
\urldef\tempurl%
\url{https://arxiv.org/abs/2112.10752}
\showURL{%
\tempurl}


\bibitem[Sch\"{o}nberger and Frahm(2016)]%
        {schoenberger2016sfm}
\bibfield{author}{\bibinfo{person}{Johannes~Lutz Sch\"{o}nberger} {and} \bibinfo{person}{Jan-Michael Frahm}.} \bibinfo{year}{2016}\natexlab{}.
\newblock \showarticletitle{Structure-from-Motion Revisited}. In \bibinfo{booktitle}{\emph{Conference on Computer Vision and Pattern Recognition (CVPR)}}.
\newblock


\bibitem[Shi et~al\mbox{.}(2024)]%
        {shi2024mvdreammultiviewdiffusion3d}
\bibfield{author}{\bibinfo{person}{Yichun Shi}, \bibinfo{person}{Peng Wang}, \bibinfo{person}{Jianglong Ye}, \bibinfo{person}{Mai Long}, \bibinfo{person}{Kejie Li}, {and} \bibinfo{person}{Xiao Yang}.} \bibinfo{year}{2024}\natexlab{}.
\newblock \bibinfo{title}{MVDream: Multi-view Diffusion for 3D Generation}.
\newblock
\showeprint[arxiv]{2308.16512}~[cs.CV]
\urldef\tempurl%
\url{https://arxiv.org/abs/2308.16512}
\showURL{%
\tempurl}


\bibitem[Shriram et~al\mbox{.}(2025)]%
        {shriram2024realmdreamer}
\bibfield{author}{\bibinfo{person}{Jaidev Shriram}, \bibinfo{person}{Alex Trevithick}, \bibinfo{person}{Lingjie Liu}, {and} \bibinfo{person}{Ravi Ramamoorthi}.} \bibinfo{year}{2025}\natexlab{}.
\newblock \showarticletitle{RealmDreamer: Text-Driven 3D Scene Generation with Inpainting and Depth Diffusion}.
\newblock \bibinfo{journal}{\emph{International Conference on 3D Vision (3DV)}}.
\newblock


\bibitem[Tang et~al\mbox{.}(2023)]%
        {Tang2023mvdiffusion}
\bibfield{author}{\bibinfo{person}{Shitao Tang}, \bibinfo{person}{Fuyang Zhang}, \bibinfo{person}{Jiacheng Chen}, \bibinfo{person}{Peng Wang}, {and} \bibinfo{person}{Yasutaka Furukawa}.} \bibinfo{year}{2023}\natexlab{}.
\newblock \showarticletitle{MVDiffusion: Enabling Holistic Multi-view Image Generation with Correspondence-Aware Diffusion}.
\newblock \bibinfo{journal}{\emph{arXiv}} (\bibinfo{year}{2023}).
\newblock


\bibitem[Tian et~al\mbox{.}(2024)]%
        {tian2024VAM}
\bibfield{author}{\bibinfo{person}{Keyu Tian}, \bibinfo{person}{Yi Jiang}, \bibinfo{person}{Zehuan Yuan}, \bibinfo{person}{Bingyue Peng}, {and} \bibinfo{person}{Liwei Wang}.} \bibinfo{year}{2024}\natexlab{}.
\newblock \showarticletitle{Visual Autoregressive Modeling: Scalable Image Generation via Next-Scale Prediction}. In \bibinfo{booktitle}{\emph{Advances in Neural Information Processing Systems}}, \bibfield{editor}{\bibinfo{person}{A.~Globerson}, \bibinfo{person}{L.~Mackey}, \bibinfo{person}{D.~Belgrave}, \bibinfo{person}{A.~Fan}, \bibinfo{person}{U.~Paquet}, \bibinfo{person}{J.~Tomczak}, {and} \bibinfo{person}{C.~Zhang}} (Eds.), Vol.~\bibinfo{volume}{37}. \bibinfo{publisher}{Curran Associates, Inc.}, \bibinfo{pages}{84839--84865}.
\newblock
\urldef\tempurl%
\url{https://proceedings.neurips.cc/paper_files/paper/2024/file/9a24e284b187f662681440ba15c416fb-Paper-Conference.pdf}
\showURL{%
\tempurl}


\bibitem[Voleti et~al\mbox{.}(2024)]%
        {voleti2024sv3dnovelmultiviewsynthesis}
\bibfield{author}{\bibinfo{person}{Vikram Voleti}, \bibinfo{person}{Chun-Han Yao}, \bibinfo{person}{Mark Boss}, \bibinfo{person}{Adam Letts}, \bibinfo{person}{David Pankratz}, \bibinfo{person}{Dmitry Tochilkin}, \bibinfo{person}{Christian Laforte}, \bibinfo{person}{Robin Rombach}, {and} \bibinfo{person}{Varun Jampani}.} \bibinfo{year}{2024}\natexlab{}.
\newblock \bibinfo{title}{SV3D: Novel Multi-view Synthesis and 3D Generation from a Single Image using Latent Video Diffusion}.
\newblock
\showeprint[arxiv]{2403.12008}~[cs.CV]
\urldef\tempurl%
\url{https://arxiv.org/abs/2403.12008}
\showURL{%
\tempurl}


\bibitem[Wallingford et~al\mbox{.}(2024)]%
        {wallingford2024image}
\bibfield{author}{\bibinfo{person}{Matthew Wallingford}, \bibinfo{person}{Anand Bhattad}, \bibinfo{person}{Aditya Kusupati}, \bibinfo{person}{Vivek Ramanujan}, \bibinfo{person}{Matt Deitke}, \bibinfo{person}{Aniruddha Kembhavi}, \bibinfo{person}{Roozbeh Mottaghi}, \bibinfo{person}{Wei-Chiu Ma}, {and} \bibinfo{person}{Ali Farhadi}.} \bibinfo{year}{2024}\natexlab{}.
\newblock \showarticletitle{From an Image to a Scene: Learning to Imagine the World from a Million 360° Videos}.
\newblock \bibinfo{journal}{\emph{Advances in Neural Information Processing Systems}}  \bibinfo{volume}{37} (\bibinfo{year}{2024}), \bibinfo{pages}{17743--17760}.
\newblock


\bibitem[Wang et~al\mbox{.}(2025)]%
        {wang2025videoscenedistillingvideodiffusion}
\bibfield{author}{\bibinfo{person}{Hanyang Wang}, \bibinfo{person}{Fangfu Liu}, \bibinfo{person}{Jiawei Chi}, {and} \bibinfo{person}{Yueqi Duan}.} \bibinfo{year}{2025}\natexlab{}.
\newblock \bibinfo{title}{VideoScene: Distilling Video Diffusion Model to Generate 3D Scenes in One Step}.
\newblock
\showeprint[arxiv]{2504.01956}~[cs.CV]
\urldef\tempurl%
\url{https://arxiv.org/abs/2504.01956}
\showURL{%
\tempurl}


\bibitem[Wang et~al\mbox{.}(2023)]%
        {wang2023pf}
\bibfield{author}{\bibinfo{person}{Peng Wang}, \bibinfo{person}{Hao Tan}, \bibinfo{person}{Sai Bi}, \bibinfo{person}{Yinghao Xu}, \bibinfo{person}{Fujun Luan}, \bibinfo{person}{Kalyan Sunkavalli}, \bibinfo{person}{Wenping Wang}, \bibinfo{person}{Zexiang Xu}, {and} \bibinfo{person}{Kai Zhang}.} \bibinfo{year}{2023}\natexlab{}.
\newblock \showarticletitle{PF-LRM: Pose-Free Large Reconstruction Model for Joint Pose and Shape Prediction}.
\newblock \bibinfo{journal}{\emph{arXiv preprint arXiv:2311.12024}} (\bibinfo{year}{2023}).
\newblock


\bibitem[Wang et~al\mbox{.}(2024)]%
        {wang2024360dvd}
\bibfield{author}{\bibinfo{person}{Qian Wang}, \bibinfo{person}{Weiqi Li}, \bibinfo{person}{Chong Mou}, \bibinfo{person}{Xinhua Cheng}, {and} \bibinfo{person}{Jian Zhang}.} \bibinfo{year}{2024}\natexlab{}.
\newblock \showarticletitle{360DVD: Controllable Panorama Video Generation with 360-Degree Video Diffusion Model}.
\newblock \bibinfo{journal}{\emph{arXiv preprint arXiv:2401.06578}} (\bibinfo{year}{2024}).
\newblock


\bibitem[Wang et~al\mbox{.}(2018)]%
        {wang2018esrgan}
\bibfield{author}{\bibinfo{person}{Xintao Wang}, \bibinfo{person}{Ke Yu}, \bibinfo{person}{Shixiang Wu}, \bibinfo{person}{Jinjin Gu}, \bibinfo{person}{Yihao Liu}, \bibinfo{person}{Chao Dong}, \bibinfo{person}{Yu Qiao}, {and} \bibinfo{person}{Chen~Change Loy}.} \bibinfo{year}{2018}\natexlab{}.
\newblock \showarticletitle{ESRGAN: Enhanced super-resolution generative adversarial networks}. In \bibinfo{booktitle}{\emph{The European Conference on Computer Vision Workshops (ECCVW)}}.
\newblock


\bibitem[Weber et~al\mbox{.}(2022)]%
        {hweberEditableIndoorLight}
\bibfield{author}{\bibinfo{person}{Henrique Weber}, \bibinfo{person}{Mathieu Garon}, {and} \bibinfo{person}{Jean-Fran{\c{c}}ois Lalonde}.} \bibinfo{year}{2022}\natexlab{}.
\newblock \showarticletitle{Editable Indoor Lighting Estimation}. In \bibinfo{booktitle}{\emph{European Conference on Computer Vision (ECCV)}}.
\newblock


\bibitem[Wei et~al\mbox{.}(2024)]%
        {wei2024meshlrm}
\bibfield{author}{\bibinfo{person}{Xinyue Wei}, \bibinfo{person}{Kai Zhang}, \bibinfo{person}{Sai Bi}, \bibinfo{person}{Hao Tan}, \bibinfo{person}{Fujun Luan}, \bibinfo{person}{Valentin Deschaintre}, \bibinfo{person}{Kalyan Sunkavalli}, \bibinfo{person}{Hao Su}, {and} \bibinfo{person}{Zexiang Xu}.} \bibinfo{year}{2024}\natexlab{}.
\newblock \showarticletitle{MeshLRM: Large Reconstruction Model for High-Quality Mesh}.
\newblock \bibinfo{journal}{\emph{arXiv preprint arXiv:2404.12385}} (\bibinfo{year}{2024}).
\newblock


\bibitem[Wu et~al\mbox{.}(2023)]%
        {wu2023qalign}
\bibfield{author}{\bibinfo{person}{Haoning Wu}, \bibinfo{person}{Zicheng Zhang}, \bibinfo{person}{Weixia Zhang}, \bibinfo{person}{Chaofeng Chen}, \bibinfo{person}{Chunyi Li}, \bibinfo{person}{Liang Liao}, \bibinfo{person}{Annan Wang}, \bibinfo{person}{Erli Zhang}, \bibinfo{person}{Wenxiu Sun}, \bibinfo{person}{Qiong Yan}, \bibinfo{person}{Xiongkuo Min}, \bibinfo{person}{Guangtai Zhai}, {and} \bibinfo{person}{Weisi Lin}.} \bibinfo{year}{2023}\natexlab{}.
\newblock \showarticletitle{Q-Align: Teaching LMMs for Visual Scoring via Discrete Text-Defined Levels}.
\newblock \bibinfo{journal}{\emph{arXiv preprint arXiv:2312.17090}} (\bibinfo{year}{2023}).
\newblock
\newblock
\shownote{Equal Contribution by Wu, Haoning and Zhang, Zicheng. Project Lead by Wu, Haoning. Corresponding Authors: Zhai, Guangtai and Lin, Weisi.}.


\bibitem[Wu et~al\mbox{.}(2025)]%
        {wu2025videoworldmodelslongterm}
\bibfield{author}{\bibinfo{person}{Tong Wu}, \bibinfo{person}{Shuai Yang}, \bibinfo{person}{Ryan Po}, \bibinfo{person}{Yinghao Xu}, \bibinfo{person}{Ziwei Liu}, \bibinfo{person}{Dahua Lin}, {and} \bibinfo{person}{Gordon Wetzstein}.} \bibinfo{year}{2025}\natexlab{}.
\newblock \bibinfo{title}{Video World Models with Long-term Spatial Memory}.
\newblock
\showeprint[arxiv]{2506.05284}~[cs.CV]
\urldef\tempurl%
\url{https://arxiv.org/abs/2506.05284}
\showURL{%
\tempurl}


\bibitem[Wu et~al\mbox{.}(2024)]%
        {wu2024recentadvances3dgaussian}
\bibfield{author}{\bibinfo{person}{Tong Wu}, \bibinfo{person}{Yu-Jie Yuan}, \bibinfo{person}{Ling-Xiao Zhang}, \bibinfo{person}{Jie Yang}, \bibinfo{person}{Yan-Pei Cao}, \bibinfo{person}{Ling-Qi Yan}, {and} \bibinfo{person}{Lin Gao}.} \bibinfo{year}{2024}\natexlab{}.
\newblock \bibinfo{title}{Recent Advances in 3D Gaussian Splatting}.
\newblock
\showeprint[arxiv]{2403.11134}~[cs.CV]
\urldef\tempurl%
\url{https://arxiv.org/abs/2403.11134}
\showURL{%
\tempurl}


\bibitem[Xie et~al\mbox{.}(2024)]%
        {xie2024lrmzerotraininglargereconstruction}
\bibfield{author}{\bibinfo{person}{Desai Xie}, \bibinfo{person}{Sai Bi}, \bibinfo{person}{Zhixin Shu}, \bibinfo{person}{Kai Zhang}, \bibinfo{person}{Zexiang Xu}, \bibinfo{person}{Yi Zhou}, \bibinfo{person}{Sören Pirk}, \bibinfo{person}{Arie Kaufman}, \bibinfo{person}{Xin Sun}, {and} \bibinfo{person}{Hao Tan}.} \bibinfo{year}{2024}\natexlab{}.
\newblock \bibinfo{title}{LRM-Zero: Training Large Reconstruction Models with Synthesized Data}.
\newblock
\showeprint[arxiv]{2406.09371}~[cs.CV]
\urldef\tempurl%
\url{https://arxiv.org/abs/2406.09371}
\showURL{%
\tempurl}


\bibitem[Yang et~al\mbox{.}(2024)]%
        {yang2024layerpano3d}
\bibfield{author}{\bibinfo{person}{Shuai Yang}, \bibinfo{person}{Jing Tan}, \bibinfo{person}{Mengchen Zhang}, \bibinfo{person}{Tong Wu}, \bibinfo{person}{Yixuan Li}, \bibinfo{person}{Gordon Wetzstein}, \bibinfo{person}{Ziwei Liu}, {and} \bibinfo{person}{Dahua Lin}.} \bibinfo{year}{2024}\natexlab{}.
\newblock \showarticletitle{LayerPano3D: Layered 3D Panorama for Hyper-Immersive Scene Generation}.
\newblock \bibinfo{journal}{\emph{arXiv preprint arXiv:2408.13252}} (\bibinfo{year}{2024}).
\newblock


\bibitem[Ye et~al\mbox{.}(2024b)]%
        {ye2024gsplatopensourcelibrarygaussian}
\bibfield{author}{\bibinfo{person}{Vickie Ye}, \bibinfo{person}{Ruilong Li}, \bibinfo{person}{Justin Kerr}, \bibinfo{person}{Matias Turkulainen}, \bibinfo{person}{Brent Yi}, \bibinfo{person}{Zhuoyang Pan}, \bibinfo{person}{Otto Seiskari}, \bibinfo{person}{Jianbo Ye}, \bibinfo{person}{Jeffrey Hu}, \bibinfo{person}{Matthew Tancik}, {and} \bibinfo{person}{Angjoo Kanazawa}.} \bibinfo{year}{2024}\natexlab{b}.
\newblock \showarticletitle{gsplat: An Open-Source Library for {Gaussian} Splatting}.
\newblock \bibinfo{journal}{\emph{arXiv preprint arXiv:2409.06765}} (\bibinfo{year}{2024}).
\newblock
\showeprint[arxiv]{2409.06765}~[cs.CV]
\urldef\tempurl%
\url{https://arxiv.org/abs/2409.06765}
\showURL{%
\tempurl}


\bibitem[Ye et~al\mbox{.}(2024a)]%
        {ye2024diffpanoscalableconsistenttext}
\bibfield{author}{\bibinfo{person}{Weicai Ye}, \bibinfo{person}{Chenhao Ji}, \bibinfo{person}{Zheng Chen}, \bibinfo{person}{Junyao Gao}, \bibinfo{person}{Xiaoshui Huang}, \bibinfo{person}{Song-Hai Zhang}, \bibinfo{person}{Wanli Ouyang}, \bibinfo{person}{Tong He}, \bibinfo{person}{Cairong Zhao}, {and} \bibinfo{person}{Guofeng Zhang}.} \bibinfo{year}{2024}\natexlab{a}.
\newblock \bibinfo{title}{DiffPano: Scalable and Consistent Text to Panorama Generation with Spherical Epipolar-Aware Diffusion}.
\newblock
\showeprint[arxiv]{2410.24203}~[cs.CV]
\urldef\tempurl%
\url{https://arxiv.org/abs/2410.24203}
\showURL{%
\tempurl}


\bibitem[Yu et~al\mbox{.}(2024a)]%
        {yu2024wonderworldinteractive3dscene}
\bibfield{author}{\bibinfo{person}{Hong-Xing Yu}, \bibinfo{person}{Haoyi Duan}, \bibinfo{person}{Charles Herrmann}, \bibinfo{person}{William~T. Freeman}, {and} \bibinfo{person}{Jiajun Wu}.} \bibinfo{year}{2024}\natexlab{a}.
\newblock \bibinfo{title}{WonderWorld: Interactive 3D Scene Generation from a Single Image}.
\newblock
\showeprint[arxiv]{2406.09394}~[cs.CV]
\urldef\tempurl%
\url{https://arxiv.org/abs/2406.09394}
\showURL{%
\tempurl}


\bibitem[Yu et~al\mbox{.}(2024b)]%
        {yu2024language}
\bibfield{author}{\bibinfo{person}{Lijun Yu}, \bibinfo{person}{Jos{\'e} Lezama}, \bibinfo{person}{Nitesh~B Gundavarapu}, \bibinfo{person}{Luca Versari}, \bibinfo{person}{Kihyuk Sohn}, \bibinfo{person}{David Minnen}, \bibinfo{person}{Yong Cheng}, \bibinfo{person}{Agrim Gupta}, \bibinfo{person}{Xiuye Gu}, \bibinfo{person}{Alexander~G Hauptmann}, {et~al\mbox{.}}} \bibinfo{year}{2024}\natexlab{b}.
\newblock \showarticletitle{LANGUAGE MODEL BEATS DIFFUSION-TOKENIZER IS KEY TO VISUAL GENERATION}. In \bibinfo{booktitle}{\emph{12th International Conference on Learning Representations, ICLR 2024}}.
\newblock


\bibitem[Yu et~al\mbox{.}(2024c)]%
        {yu2024viewcraftertamingvideodiffusion}
\bibfield{author}{\bibinfo{person}{Wangbo Yu}, \bibinfo{person}{Jinbo Xing}, \bibinfo{person}{Li Yuan}, \bibinfo{person}{Wenbo Hu}, \bibinfo{person}{Xiaoyu Li}, \bibinfo{person}{Zhipeng Huang}, \bibinfo{person}{Xiangjun Gao}, \bibinfo{person}{Tien-Tsin Wong}, \bibinfo{person}{Ying Shan}, {and} \bibinfo{person}{Yonghong Tian}.} \bibinfo{year}{2024}\natexlab{c}.
\newblock \bibinfo{title}{ViewCrafter: Taming Video Diffusion Models for High-fidelity Novel View Synthesis}.
\newblock
\showeprint[arxiv]{2409.02048}~[cs.CV]
\urldef\tempurl%
\url{https://arxiv.org/abs/2409.02048}
\showURL{%
\tempurl}


\bibitem[Zhan et~al\mbox{.}(2021)]%
        {zhan2021emlight}
\bibfield{author}{\bibinfo{person}{Fangneng Zhan}, \bibinfo{person}{Changgong Zhang}, \bibinfo{person}{Yingchen Yu}, \bibinfo{person}{Yuan Chang}, \bibinfo{person}{Shijian Lu}, \bibinfo{person}{Feiying Ma}, {and} \bibinfo{person}{Xuansong Xie}.} \bibinfo{year}{2021}\natexlab{}.
\newblock \showarticletitle{EMLight: Lighting Estimation via Spherical Distribution Approximation}. In \bibinfo{booktitle}{\emph{Proceedings of the AAAI Conference on Artificial Intelligence}}.
\newblock


\bibitem[Zhang et~al\mbox{.}(2024b)]%
        {panfusion2024}
\bibfield{author}{\bibinfo{person}{Cheng Zhang}, \bibinfo{person}{Qianyi Wu}, \bibinfo{person}{Camilo Cruz~Gambardella}, \bibinfo{person}{Xiaoshui Huang}, \bibinfo{person}{Dinh Phung}, \bibinfo{person}{Wanli Ouyang}, {and} \bibinfo{person}{Jianfei Cai}.} \bibinfo{year}{2024}\natexlab{b}.
\newblock \showarticletitle{Taming Stable Diffusion for Text to 360◦ Panorama Image Generation}. In \bibinfo{booktitle}{\emph{Proceedings of the IEEE/CVF Conference on Computer Vision and Pattern Recognition}}.
\newblock


\bibitem[Zhang et~al\mbox{.}(2024a)]%
        {gslrm2024}
\bibfield{author}{\bibinfo{person}{Kai Zhang}, \bibinfo{person}{Sai Bi}, \bibinfo{person}{Hao Tan}, \bibinfo{person}{Yuanbo Xiangli}, \bibinfo{person}{Nanxuan Zhao}, \bibinfo{person}{Kalyan Sunkavalli}, {and} \bibinfo{person}{Zexiang Xu}.} \bibinfo{year}{2024}\natexlab{a}.
\newblock \showarticletitle{GS-LRM: Large Reconstruction Model for 3D Gaussian Splatting}.
\newblock \bibinfo{journal}{\emph{European Conference on Computer Vision}} (\bibinfo{year}{2024}).
\newblock


\bibitem[Zheng et~al\mbox{.}(2024)]%
        {opensora}
\bibfield{author}{\bibinfo{person}{Zangwei Zheng}, \bibinfo{person}{Xiangyu Peng}, \bibinfo{person}{Tianji Yang}, \bibinfo{person}{Chenhui Shen}, \bibinfo{person}{Shenggui Li}, \bibinfo{person}{Hongxin Liu}, \bibinfo{person}{Yukun Zhou}, \bibinfo{person}{Tianyi Li}, {and} \bibinfo{person}{Yang You}.} \bibinfo{year}{2024}\natexlab{}.
\newblock \bibinfo{booktitle}{\emph{Open-Sora: Democratizing Efficient Video Production for All}}.
\newblock
\urldef\tempurl%
\url{https://github.com/hpcaitech/Open-Sora}
\showURL{%
\tempurl}


\bibitem[Zhou et~al\mbox{.}(2024)]%
        {zhou2025dreamscene360}
\bibfield{author}{\bibinfo{person}{Shijie Zhou}, \bibinfo{person}{Zhiwen Fan}, \bibinfo{person}{Dejia Xu}, \bibinfo{person}{Haoran Chang}, \bibinfo{person}{Pradyumna Chari}, \bibinfo{person}{Tejas Bharadwaj}, \bibinfo{person}{Suya You}, \bibinfo{person}{Zhangyang Wang}, {and} \bibinfo{person}{Achuta Kadambi}.} \bibinfo{year}{2024}\natexlab{}.
\newblock \showarticletitle{DreamScene360: Unconstrained Text-to-3D Scene Generation with Panoramic Gaussian Splatting}. In \bibinfo{booktitle}{\emph{Computer Vision – ECCV 2024: 18th European Conference, Milan, Italy, September 29–October 4, 2024, Proceedings, Part VI}} (Milan, Italy). \bibinfo{publisher}{Springer-Verlag}, \bibinfo{address}{Berlin, Heidelberg}, \bibinfo{pages}{324–342}.
\newblock
\showISBNx{978-3-031-72657-6}
\href{https://doi.org/10.1007/978-3-031-72658-3_19}{doi:\nolinkurl{10.1007/978-3-031-72658-3_19}}


\bibitem[Ziwen et~al\mbox{.}(2024)]%
        {ziwen2024llrm}
\bibfield{author}{\bibinfo{person}{Chen Ziwen}, \bibinfo{person}{Hao Tan}, \bibinfo{person}{Kai Zhang}, \bibinfo{person}{Sai Bi}, \bibinfo{person}{Fujun Luan}, \bibinfo{person}{Yicong Hong}, \bibinfo{person}{Li Fuxin}, {and} \bibinfo{person}{Zexiang Xu}.} \bibinfo{year}{2024}\natexlab{}.
\newblock \showarticletitle{Long-LRM: Long-sequence Large Reconstruction Model for Wide-coverage Gaussian Splats}.
\newblock \bibinfo{journal}{\emph{arXiv preprint 2410.12781}} (\bibinfo{year}{2024}).
\newblock


\end{thebibliography}

\end{document}